**Pressure engineering of the excitonic moiré-potential landscape and exciton-charge interplay in semiconductor moiré superlattices**

Wenbing Li[1,2]†, Tianhao Yue[1,2]†, Wenjian Su[1,2], Zhexu Shan[1,2], Chen Xu[1,2], Jichen Zhou[1,2], Kenji Watanabe[3], Takashi Taniguchi[4], Liguo Ma[1,2], Yuanfeng Xu[1,5], Shiyao Zhu[1,2,6,7,8], Yang Gao[9,10,§] and Yanhao Tang[1,2,*]

[1] School of Physics, Zhejiang University, Hangzhou 310027, China

[2] Zhejiang Key Laboratory of Micro-Nano Quantum Chips and Quantum Control, Zhejiang University, Hangzhou 310027, China

[3] Research Center for Functional Materials, National Institute for Materials Science, 1-1 Namiki, 305-0044 Tsukuba, Japan

[4] International Center for Materials Nanoarchitectonics, National Institute for Materials Science, 1-1 Namiki, 305-0044 Tsukuba, Japan

[5] Center for Correlated Matter, Zhejiang University, Hangzhou 310027, China

[6] State Key Laboratory for Extreme Photonics and Instrumentation, Zhejiang University, Hangzhou 310027, China

[7] College of Optical Science and Engineering, Zhejiang University, Hangzhou 310027, China

[8] Hefei National Laboratory, Hefei, China

[9] Center for X-Mechanics, Department of Engineering Mechanics, Zhejiang University, Hangzhou 310027, China

[10] Key Laboratory of Soft Machines and Smart Devices of Zhejiang Province, Zhejiang University, Hangzhou 310027, China

†These authors contributed equally to this work.

§Contact author: ygao96@zju.edu.cn

*Contact author: yanhaotc@zju.edu.cn

**Abstract:**

Semiconductor moiré superlattices, owing to strongly bound excitons and pronounced electronic correlation, have emerged as an intriguing solid platform for exploring correlated bosonic phenomena, where the moiré-potential landscape plays a central role. Hydrostatic pressure allows continuous, in-situ tuning of this landscape, but remains underexplored because of experimental challenges. Here we integrate a diamond-anvil pressure cell with high-quality dual-gated $WSe_2/WS_2$ twisted bilayers and perform cryogenic optical spectroscopy. We observe distinct pressure evolutions of moiré intralayer excitons, which reveal drastic modulation of the excitonic moiré-potential landscape captured by a continuum model and suggest the capability of engineering the exciton-band topology. Moreover, we demonstrate efficient pressure control over the interplay between charges and both intralayer and interlayer excitons: the intralayer charge-transfer exciton becomes less sensitive to doping, likely due to reduced charge-transfer character; in contrast, the interlayer exciton exhibits enhanced exciton–charge repulsion. Our results establish hydrostatic pressure as a powerful tuning knob for engineering bosonic quasiparticles and Bose-Fermi interactions in moiré materials.

*Introduction*—Semiconductor transition-metal-dichalcogenides (TMD) moiré superlattices, owing to exotic interplay between flat moiré bands and Coulomb interactions, harbor plenty of correlated fermionic phases [1–13]; in addition, they host strongly bound electron-hole pairs––excitons [14]—providing a solid platform for exploring correlated bosonic physics, such as excitonic Mott insulator [15–17], dipolar excitonic insulator [18,19], exciton-enabled control of magnetism [20] and Chern insulators [21–23], and exotic low-lying excitations [24]. The excitonic moiré-potential landscape—not only the potential amplitude but also its spatial distribution—plays a crucial role underlying these emergent phenomena: it is a key ingredient that determines the exciton-band topology [25] and the exciton spatial distribution (e.g., Wannier [25] versus charge-transfer [26] type), and hence influences the exciton-charge interplay that are crucial for Bose-Fermi mixtures [15,20–23,27–29]. However, the moiré-potential landscape is generally fixed after device fabrication; while it can be tuned by varying the twist angle, this approach requires extensive device fabrication and is often hindered by poorly controlled lattice reconstruction [30].

Hydrostatic pressure enables in-situ and continuous tuning of the moiré potential [31–35]. It can enhance the potential magnitude of TMD moiré materials, as revealed by moiré excitons [33] and phonons [34], and stabilize Chern insulator [36,37]; it has also been predicted to switch the underlying Hamiltonian [38]. However, due to the technical challenges of integrating pressure with both electrostatic control and optical spectroscopy while maintaining high device quality, the pressure modulation of many-body interactions, particularly the exciton-charge interplays, remains largely unexplored; furthermore, how the excitonic moiré-potential landscape evolves under pressure remains elusive.

In this Letter, we integrate a diamond-anvil pressure cell (DAC) with dual-gated twisted $WSe_2/WS_2$ bilayers and perform low-temperature optical spectroscopy. We demonstrate efficient pressure tuning of both the excitonic moiré-potential landscape and exciton-charge interplay. Specifically, moiré intralayer excitons exhibit distinct pressure dependences, accompanied by the emergence of a higher-energy resonance; these observations are well captured by a phenomenological model, which indicates that pressure enhances the potential amplitude and steepens its profile, and further predicts a feasible pressure-driven topological transition between exciton bands with Chern numbers 0 and $\pm 1$. For exciton-charge interplay, pressure reduces the doping-induced redshift of the intralayer charge-transfer exciton, indicating a diminished charge-transfer character; by contrast, pressure enhances the energy shift of the interlayer exciton (IX) at filling of one (one electron per moiré unit cell), reflecting an increase in the IX-electron repulsion.

*Device and hydrostatic pressure*—Fig. 1(a) illustrate the DAC integrating a dual-gated heterostructure that sits on the diamond anvil culet with prepatterned electrodes. The device schematic [Fig. 1(b)] depicts a twisted $WSe_2/WS_2$ bilayer encapsulated in hBN, grounded through a graphite electrode, and dual-gated by top and bottom graphite layers. The top graphite is Au-coated to minimize the optical interference in reflectance measurements. See details in Supplemental Material [39].

We have investigated two $WSe_2/WS_2$ bilayers: a nearly-60°-twisted (angle-aligned) bilayer forming a ~8 nm moiré superlattice, and a nearly 20°-twisted (misaligned) control bilayer without moiré effects. The bilayer has a type-II band alignment [Fig. 1(c)], in which intra- and interlayer excitons can be probed by reflectance contrast (*RC*) and photoluminescence (PL) spectroscopies, respectively. At a low pressure of 0.5 GPa, the moiré bilayer shows spectral features consistent with previous ambient-pressure studies [2,40,41]. Three moiré intralayer exciton resonances ($X_I$, $X_{II}$, $X_{III}$) of $WSe_2$ emerge at charge neutrality [Fig. 1(d)]; the two higher-energy ones ($X_{II}$, $X_{III}$) originate from the finite-momentum excitons hybridized with the zero-momentum state via moiré Umklapp scattering [25]. Both intra- and interlayer excitons show abrupt spectral changes at integer fillings, arising from the formation of insulating states (Fig. 1(d),(e); see filling assignments in Fig. S1 [39]), which reflect the exciton-charge interplay that will be discussed later. In contrast, the misaligned bilayer (Fig. S2 [39]) shows only a single intralayer exciton resonance ($X_A$) of $WSe_2$, besides the Rydberg state ($X_{2s}$), with gradual doping behavior similar to that of monolayer [42].

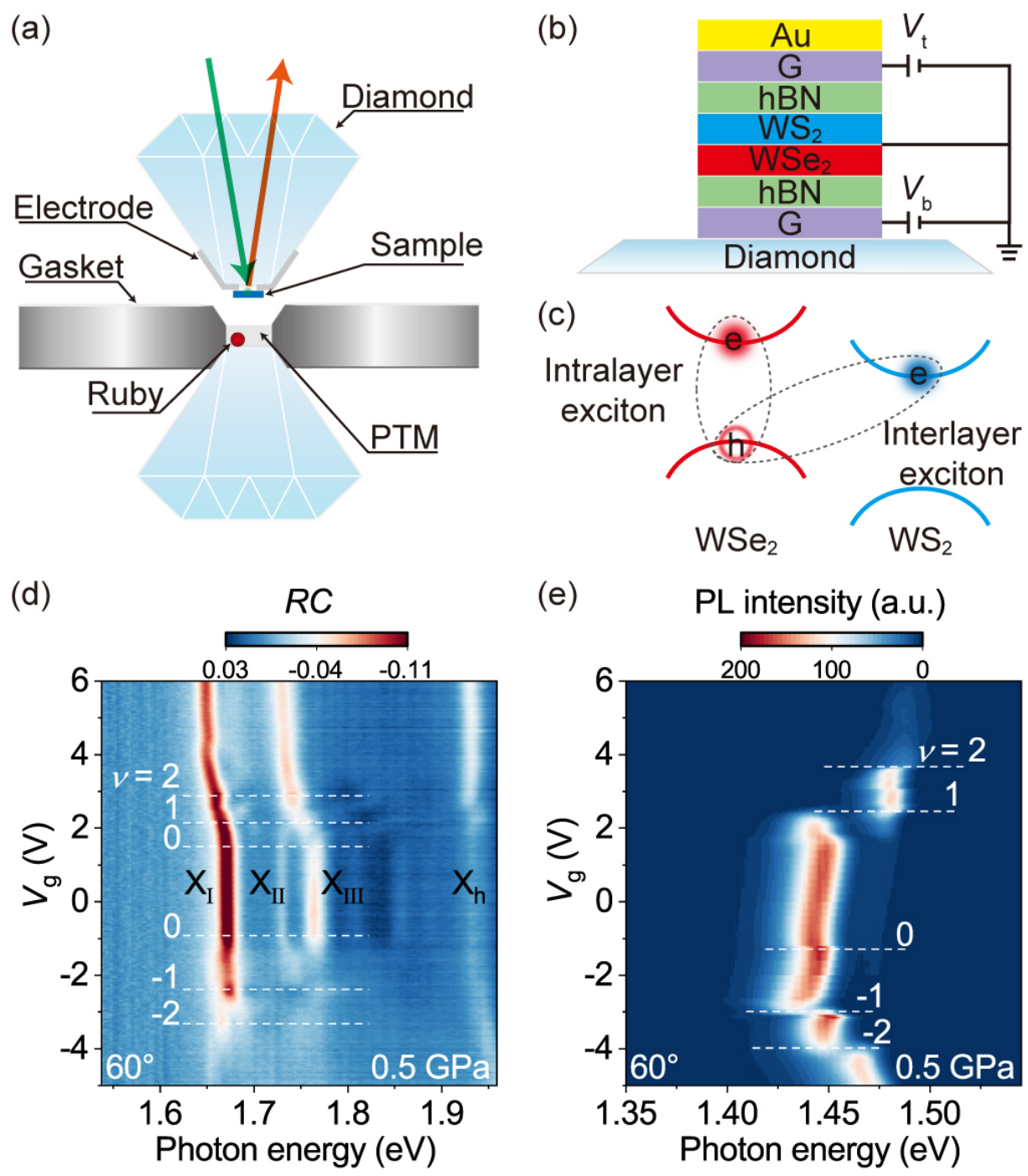

FIG. 1. Pressure setup and optical characterization of twisted $WSe_2/WS_2$ bilayers. (a) Schematics of the DAC setup. (b) The horizontal view of the device. (c) The type-II band alignment of twisted $WSe_2/WS_2$ bilayers. (d),(e) Doping-dependent *RC* (d) and photoluminescence (e) spectra of the 60°-twisted bilayer at 0.5 GPa. $X_I$, $X_{II}$, $X_{III}$ and $X_{IV}$ denote moiré intralayer excitons of $WSe_2$; $X_h$ denote the spatially hybridized exciton. The dashed lines denote integer fillings ($\nu$).

*Pressure tuning moiré potential*—Moiré excitons provide the fingerprints of the excitonic moiré-potential landscape. We explore the pressure dependence of $WSe_2$ moiré intralayer excitons by *RC* measurements as shown in Fig. 2(b); the bilayer is kept at charge neutral, and a smooth background has been subtracted from the spectrum to better resolve the resonances (see Fig. S3 [39]). See intralayer excitons in $WS_2$ spectral range in Fig. S4 and Sect.6 of Supplemental Material [39]. As the pressure increases to 3.6 GPa, the moiré intralayer excitons exhibit distinct pressure responses: $X_{\mathrm{I}}$ redshifts by 85 meV with reduced spectral weight; $X_{\mathrm{III}}$ redshifts by 43 meV with enhanced spectral weight; $X_{\mathrm{II}}$ shows a slight blueshift of about 10 meV at 0.6 GPa with decreasing spectral weight, and becomes unresolvable above 1.2 GPa (see exciton assignment in Fig. S5 [39]). Notably, a new higher-energy resonance, $X_{\mathrm{IV}}$, emerges above 1.8 GPa and gains spectral weight with increasing pressure. See details in Fig. S6 [39]. In contrast, $X_{\mathrm{A}}$ of the misaligned bilayer blueshifts by 27 meV up to 4.2 GPa [Fig. 2(a)]. The pressure dependence of moiré intralayer excitons has been reproduced at another location on the same device (Fig. S7 [39]). Data were primarily acquired during pressurization, with select measurements during depressurization (marked by asterisks); this confirms sample reversibility, as evidenced by the similar spectra at 0.6 and 0.5 GPa. Moreover, the linewidth of moiré intralayer excitons—particularly $X_{\mathrm{I}}$—broadens significantly above 3.6 GPa (Fig. S6 [39]), indicating severe pressure-induced inhomogeneity and hence is not the focus of this work.

To elucidate the pressure dependence of moiré intralayer excitons, we adopt a phenomenological model from ref. [25], which captures the evolution of the excitonic moiré-potential landscape by a spatially varying potential $\Delta(\boldsymbol{r})$. The model imposes $\Delta(\boldsymbol{r})$ on the Hamiltonian of the center-of-mass motion A exciton of monolayer $WSe_2$, yielding the moiré exciton bands and optical conductivity (Fig. S8 [39]). $\Delta(\boldsymbol{r})$ is approximated by the first harmonic expansion, $\sum_{j=1}^{6} V\exp(i\boldsymbol{b_j}\cdot\boldsymbol{r}+i\Psi_j)$, where $\boldsymbol{b_j}$ are reciprocal moiré lattice vectors, the amplitude $V$ and the phase $\Psi_j\equiv(-1)^j\Psi$ are only simulating parameters. Other parameters are held fixed across all pressures, such as the effective exciton mass and exchange interaction (see the justification in Sect. 5 of Supplemental Material [39]). For each pressure, an optimal parameter regime of $V$ and $\Psi$ are determined by matching four spectral features: the energy separation $E_{\mathrm{III}}-E_{\mathrm{I}}$ and three spectral weight ratios $A_i/A_{\mathrm{I}}$ $(i=\mathrm{II},\mathrm{III},\mathrm{IV})$, as summarized in Fig. 2(e); see details in Sect. 5 of Supplemental Material [39]. The corresponding theoretical spectra [Fig. 2(c)], using $(V,\ \Psi)$ from the representative centers in Fig. 2(e) and Fig. S9 [39], show good agreement with experiment in both the spectral weight ratios and energy separations [Fig. 2(d)]. Notably, several closely spaced theoretical peaks [enclosed by the dashed circle in Fig. 2(c)] are attributed to $X_{\mathrm{IV}}$; they originate from multiple $\gamma$-point states of higher-energy moiré bands but are not separated experimentally, likely due to linewidth broadening (see details in Fig. S8 and Sect. 7 of Supplemental Material [39]).

The agreement between the simulation and experiment reveals a pressure-induced increase in both $V$ and $\Psi$ of the excitonic moiré potential [Fig. 2(e)]; such evolution not only enhances the peak-to-peak potential amplitude, but also steepens the potential profile near its minima [Fig. 2(f),(g)]. More importantly, the pressure evolution suggests a feasible means of realizing a topological transition of the exciton band. Exciton band with finite Chern numbers can be generated by combining exchange interaction, an external magnetic field, and the moiré potential, as proposed in ref. [25]. Following this approach [25], we calculate the Chern-number phase diagram as a function of $V$ and $\Psi$, using a fixed Zeeman splitting of 1.5 meV accessible experimentally [2], as shown in Fig. 2(e). The diagram exhibits two distinct regions with Chern number of +1 and 0, and the pressure-driven evolution of the moiré potential implies a transition between these regions. See details in Fig. S10 and Sect. 5 of Supplemental Material [39]. Nevertheless, we have to admit more sophisticated calculations, such as density functional theory [26], are required to further understand the pressure effects on the moiré potential and excitons.

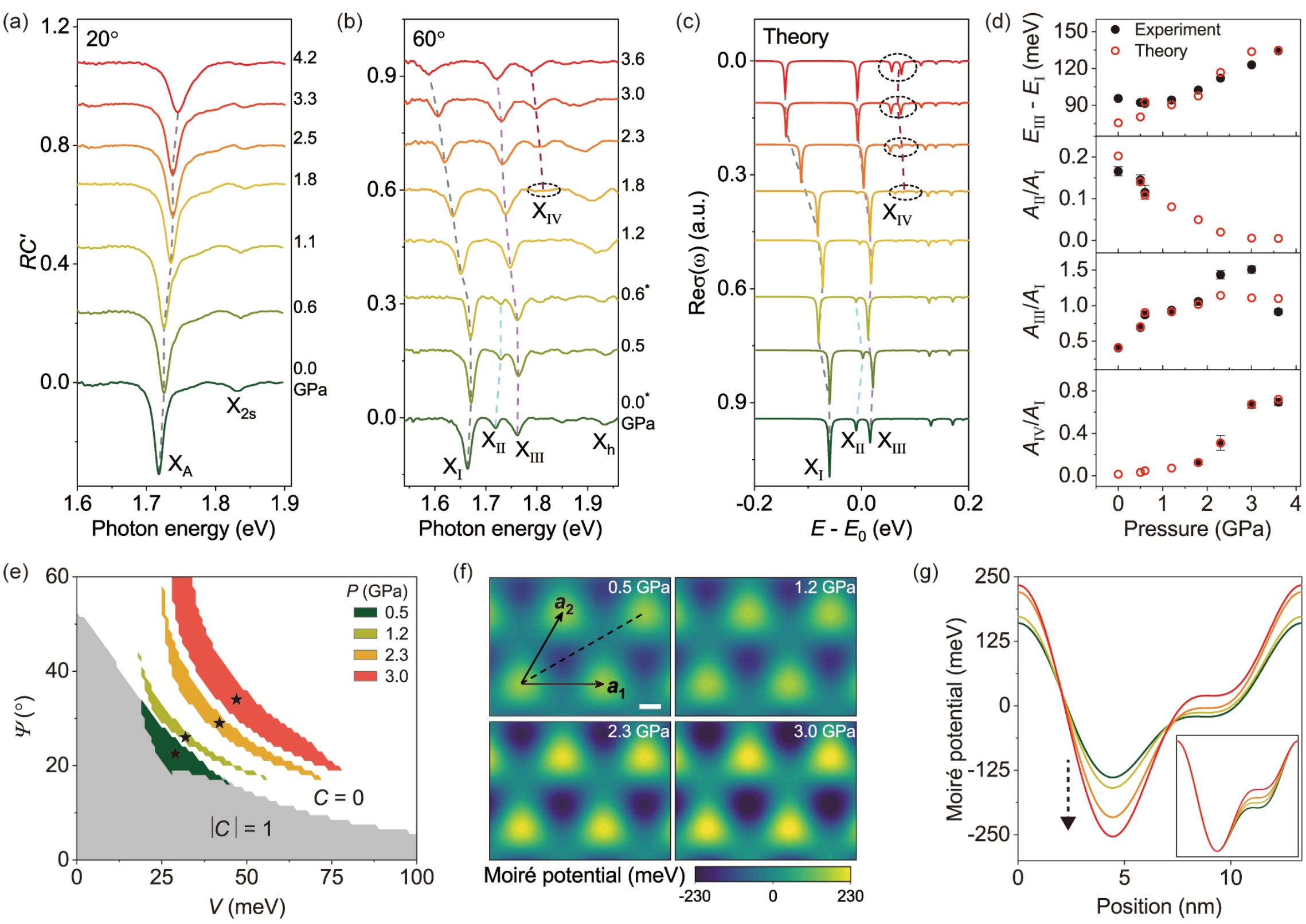

FIG. 2. Pressure-tuned moiré intralayer excitons and continuum model simulations. (a),(b) Background-subtracted $RC$ spectra of the 20°- (a) and 60°-twisted (b) bilayers at charge neutrality as a function of pressure. Asterisks indicate data acquired during decompression. (c) Simulated real-part optical conductivity spectra. Spectra are vertically offset for clarity. (d) Comparison of spectral features between the experiment and simulation. $E_i$ and $A_i$ denote the energy and spectral weight of the moiré intralayer excitons $X_i$ (i=I–IV), respectively. (e) Parameter regimes of the moiré potential at selected pressures, overlaid with the regions with Chern numbers $|C|$=1 and $C$=0. Stars mark the representative center of each parameter regime. Chern numbers are calculated under a fixed Zeeman field of 1.5 meV. (f),(g) Pressure evolution of the simulated moiré-potential landscape (f), and the corresponding profile linecuts along the dashed line (g). The dashed arrow indicates the direction of pressure increasing. The inset is normalized potential profiles. Parameters $V$ and $\Psi$ used in (c) and (f) are taken from the stars in (e) and . S9 [39]. Scale bar, 2 nm.

*Pressure tuning exciton-charge interplays*—Tuning the moiré-potential landscape modifies the spatial distributions of both excitons and charges, thereby enabling control of their mutual interactions. We first explore the pressure dependence of the interaction between electrons and intralayer excitons, where the two completely reside in different layers [Fig. 3(g)]. Fig. 3(a)-(c) show the electron-doping dependence of intralayer exciton resonances at selected pressures (see more pressures in Fig. S11 [39]). We focus on $X_{I}$ and $X_{III}$, which retain substantial oscillator strength across the relevant doping and pressure ranges, and extract their doping-dependent peak energies in Fig. 3(d),(e). We can see that, $X_{III}$ exhibits a doping-induced redshift that is strongly suppressed under pressure, in contrast to $X_{I}$, which shows nearly pressure-independent redshifts. This contrast is more clearly revealed by the energy shift between fillings of zero and two, shown in Fig. 3(f).

The doping-induced redshift of intralayer excitons arises from the Coulomb screening by electrons in the adjacent layer [Fig. 3(g)]. Specifically, this screening reduces both the quasiparticle bandgap and the exciton binding energy, which have opposing effects on the exciton energy. The net redshift of intralayer excitons evidences that the quasiparticle bandgap reduction dominates, which can be understood by an image-charge model [45] (see details in Fig. S12 and Sect.8 of Supplemental Material [39]); similar phenomena have also been observed in various two-dimensional(2D) semiconductors [43,44]. More interestingly, the reduced redshift of $X_{III}$ upon pressure suggests a decrease in its electron-hole separation based on the image-charge model [Fig. S12(b)]; this implies an evolution from a charge-transfer exciton at ambient pressure (the electron and hole are spatially separated, as reported previously[26]) to a more tightly bound exciton at higher pressure, as illustrated in [Fig. 3(g)].

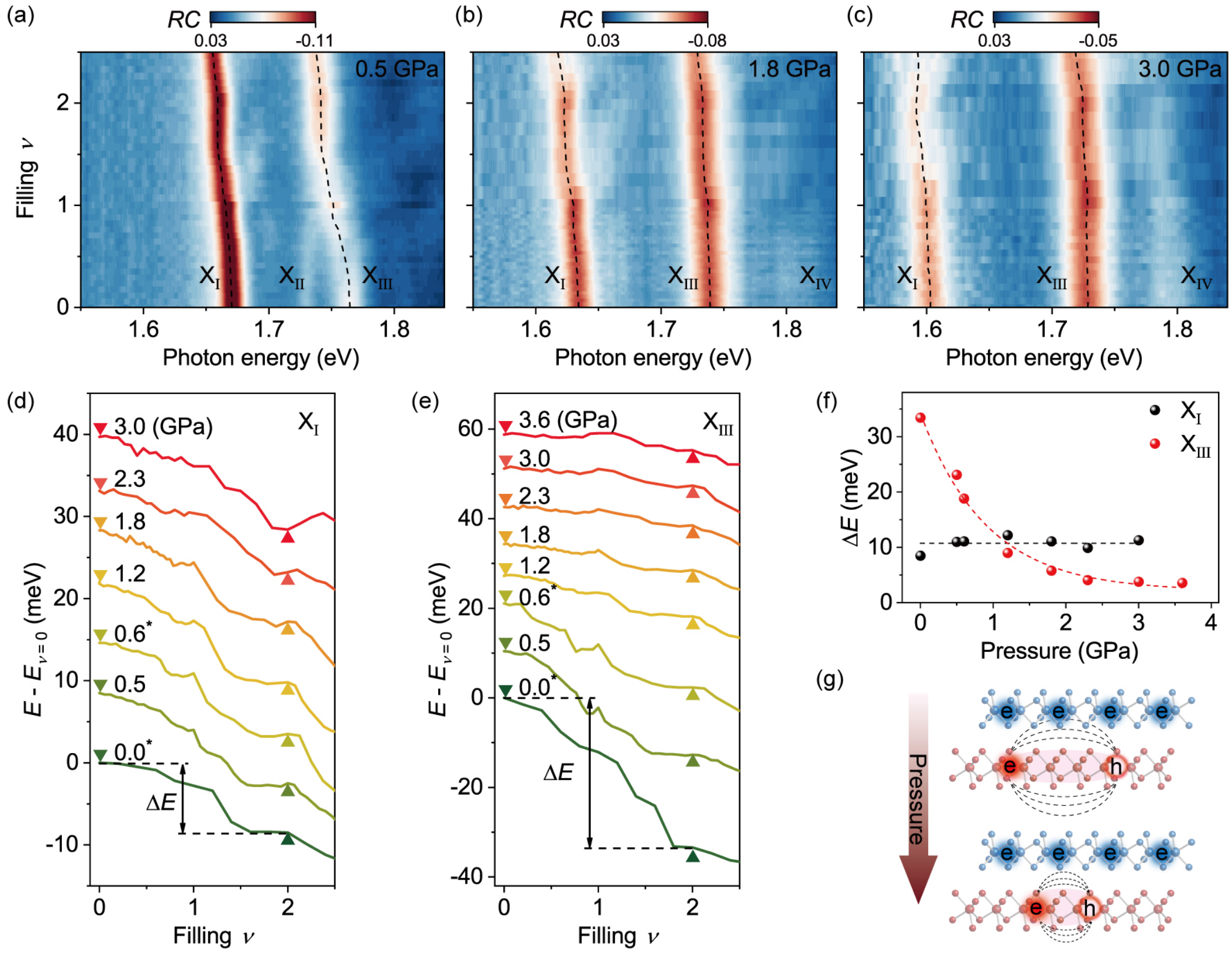

FIG. 3. Pressure-modulated interplay between electrons and moiré intralayer excitons. (a)-(c) Doping-dependent *RC* spectra of moiré intralayer excitons at 0.5 GPa (a), 1.8 GPa (b), and 3.0 GPa (c). Dashed curves trace the energy shifts of $X_{\mathrm{I}}$ and $X_{\mathrm{III}}$. (d),(e) Energy shifts of $X_{\mathrm{I}}$ (d) and $X_{\mathrm{III}}$ (e) as a function of filling, relative to the peak energy at charge neutrality ($E_{\nu=0}$). $\Delta E$ denotes the energy shift between the fillings of zero and two. Curves are vertically displaced for clarity. (f) Pressure evolution of $\Delta E$ for $X_{\mathrm{I}}$ and $X_{\mathrm{III}}$, extracted from (d) and (e), respectively. (g) Schematic illustrating the suppression of the excitonic charge-transfer character with increasing pressure.

We further explore the pressure dependence of the interaction between electrons and the IXs, where the doped electron and the electron constituent of the IX reside in the same layer [Fig. 4(e)]. Fig. 4(a),(b),(c) show the doping dependence of the IX PL resonance at selected pressures (see more pressures in Fig. S13 [39]). The IX exhibits pronounced energy shift at filling of one, denoted by $U_{\mathrm{e-IX}}$, which arise primarily from the Coulomb repulsion between the same-type carriers [17,27] as illustrated in Fig. 4(e). With increasing pressure, $U_{\mathrm{e-IX}}$ increases from 38 meV to 53 meV [Fig. 4(d)]. See details in Fig. S14 [39]. The enhancement of $U_{\mathrm{e-IX}}$ can be attributed to the pressure-induced deepening of the moiré potential: stronger confinement localizes both the IX and the electron at moiré sites, which increase their wavefunction overlap and thereby enhance the Coulomb repulsion. The IX-hole interaction exhibits a similar pressure-induced enhancement (Fig. S15 [39]). Moreover, at charge neutrality, the IX exhibits a redshift of up to 80 meV and pronounced PL suppression as pressure increases to 3.6 GPa [Fig. S16]. This redshift may arise from the pressure-enhanced moiré potential, which strengthens the hybridization between zero-momentum and momentum-folded IX states and thereby lowers the energy of the lowest IX state. Meanwhile, the PL suppression may result from enhanced K-to-Λ electron scattering as the valley energy separation decreases [46], or from increased nonradiative recombination due to pressure-induced disorder.

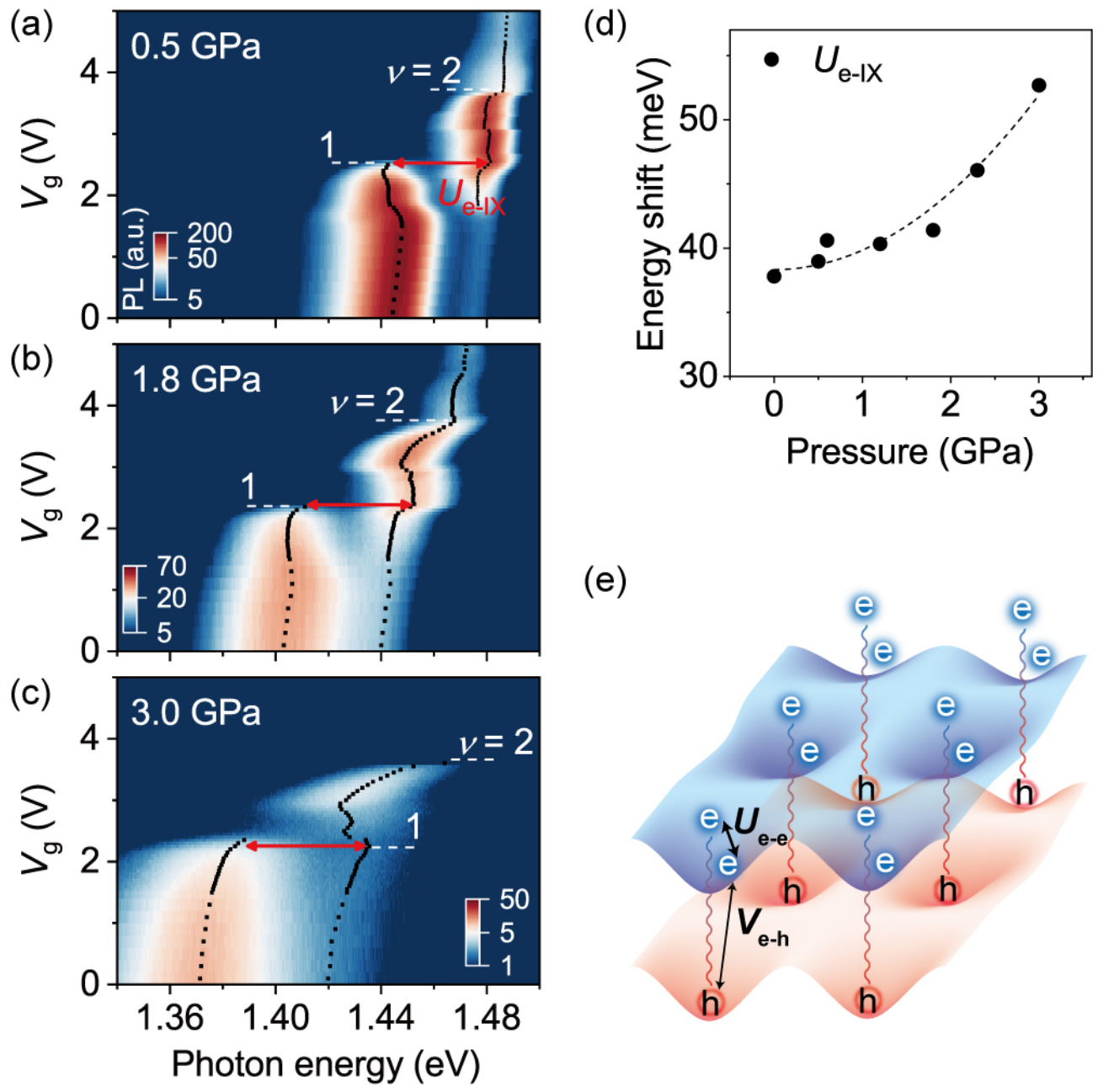

FIG. 4. Pressure-modulated interplay between doped electrons and interlayer excitons. (a)-(c) Doping-dependent PL spectra at 0.5 GPa (a) , 1.8 GPa (b) and 3.0 GPa(c). Dashed curves trace the IX peak energy. Red arrows denote the energy shifts at filling of one, denoted as $U_{e\text{-}IX}$. (d) Pressure evolution of $U_{e\text{-}IX}$. (e) Schematic of the physics mechanism governing $U_{e\text{-}IX}$.

*Discussion and conclusion*—In summary, we have demonstrated hydrostatic pressure as a powerful and in-situ tuning knob to engineer moiré excitons and their many-body interplay with charges in TMD superlattices; through a continuum model, we quantitatively estimate how the excitonic moiré-potential landscape evolves with the pressure, which enables topology engineering of moiré exciton band. Our study expands the parameter space for exploring correlated bosonic phenomena in moiré materials. Notably, a finite Bose-Fermi interaction can stabilize a supersolid phase [28,29]—a coherent bosonic fluid modulated by underlying fermionic order; by increasing such interaction, the supersolid transitions to a Bose-Fermi dual Mott insulator [29]. Future spatiotemporally resolved measurements with the pressure tuning, particularly those probing exciton diffusion [47,48], can be employed for revealing these exotic phases.

*Acknowledgments*—We thank Prof. Hongli Guo, Prof. Kin Fai Mak, Prof. Jie Shan for helpful discussions, Chenshuo Ye for assistance with the preliminary continuum model simulation. This work was supported by the National Key R&D Program of China (Grants No. 2022YFA1405400 and 2022YFA1402400), the National Natural Science Foundation of China (Grants No. 12274365 and 12550401), Zhejiang Provincial Natural Science Foundation of China (Grant No. LR24A040001 and LR26A020002), the Fundamental Research Funds for the Central Universities (Grant No. K20250213), Zhejiang Provincial Natural Science Foundation of China (Grant No. LR26A020002), Open Research Fund of the Pulsed High Magnetic Field Facility (Grant No. WHMFC2024009) of Huazhong University of Science and Technology. K.W. and T.T. acknowledge support from the JSPS KAKENHI (Grants No. 20H00354 and 23H02052) and World Premier International Research Center Initiative (WPI), MEXT, Japan. In addition, we appreciate the device fabrication support from the ZJU Micro-Nano Fabrication Center in Zhejiang University.

*Author contributions*—Y.T. conceived the scientific objectives and designed the experiments. Y.T. and Y.G. supervised the project. T.Y., W.L. and W.S. fabricated the devices with the assistance of Z.S, C.X. and J.Z.. W.L. and T.Y. developed the DAC pressure technique under supervision of Y.G. and Y.T.. W.L. and T.Y. performed optical measurements and continuum model simulation. K.W. and T.T. grew the hBN crystals. Y.T. and W.L. wrote the manuscript with the input from all the authors.

The authors declare no competing interests.

*Data availability*—The data that support the plots within this paper, and other findings of this study, are available from the corresponding authors upon reasonable request.

# Supplemental Materials for

## Pressure engineering of the excitonic moiré-potential landscape and exciton-charge interplay in semiconductor moiré superlattices

Wenbing Li *et al.*

*Corresponding author. Email: ygao96@zju.edu.cn; yanhaotc@zju.edu.cn

**The PDF file includes:**

Supplementary Text
Figs. S1 to S23

## 1. Device fabrication

The dual-gated $WSe_2/WS_2$ heterostructures were fabricated using a dry-transfer method [49]. Atomically thin flakes were exfoliated onto $SiO_2$-Si substrates, and identified with an optical microscope. A polymer stamp—a polycarbonate film on a polypropylene-carbonate-coated polydimethylsiloxane block—was employed to sequentially pick up the flakes in a nitrogen-filled glovebox. The assembled stack was then released onto the culet of a diamond anvil with pre-patterned Ti/Au electrodes. Finally, a reflective layer (10 nm Ti /90 nm Au) was deposited onto the devices using a shadow mask to eliminate the optical interference in reflectance spectroscopy. Prior to stacking, the crystal axes of monolayer $WSe_2$ and $WS_2$ were determined using polarization-resolved second-harmonic generation (SHG). The final twist angles of 0° and 60° were further distinguished by the doping dependence of the *RC* spectrum [41]. See the optical image of the device and SHG characterization in Fig. S17.

## 2. Diamond anvil cell assembly and pressure calibration

A symmetric DAC with two 400-μm-diameter culets (one hosting the device) was utilized to generate hydrostatic pressures up to ~6 GPa. The diamond anvils are type-IIa to minimize the background noise. A 250-μm-thick stainless steel gasket (T301), covered with cubic boron nitride for electrical insulation, was pre-indented and drilled to create a 200-μm-diameter sample chamber. A 4:1 methanol/ethanol mixture served as the pressure-transmitting medium (PTM) to ensure good hydrostatic conditions. Pt foils (4 μm thick) were utilized to connect the pre-patterned electrodes on the diamond culet to the outside. Ruby spheres were loaded into the sample chamber for in-situ pressure calibration.

A full pressure cycle consisted of pressurization followed by depressurization [Fig. S19(c)]. At each pressure point, the DAC was adjusted at ambient temperature and then mounted on the cold finger of a closed-cycle cryostat to reach a base temperature of ~4 K (monitored by a thermosensor attached to the DAC). The actual pressure was calibrated at 4 K using ruby fluorescence, following the method in ref. [50]. Quasi-hydrostatic conditions were maintained throughout all pressures, as evidenced by the negligible broadening of the ruby $R_1$ line emission [Fig. S19(a),(b)]. The moiré device exhibits reasonable spectral reversibility after a full pressure cycle based on the *RC* (Fig. S6), PL (Fig. S16) and Raman measurements (Fig. S18), indicating negligible pressure-induced sample damage. To be noted, pronounced spectral broadening was observed above 3.0 GPa for both the $WSe_2$ $A_{1g}$ phonon mode (Fig. S18) and the inter- and intralayer moiré excitons (Fig. S6, Fig. S16); this indicates pressure-induced inhomogeneity, which likely arises from the solidification of the PTM at cryogenic temperatures and hence degrades hydrostaticity to some extent. A PTM with better hydrostatic performance, such as helium [51], might solve this issue.

## 3. Optical measurements

For *RC* measurements, a broadband tungsten-halogen lamp was used as the light source. The output of the lamp was focused into a single-mode fiber, collimated by a 20× objective, and focused onto the sample with a spot size of about 2 μm using a 50× objective (N.A.=0.5). The reflected light was collected through the same objective, and directed into a monochromator equipped with a liquid-nitrogen-cooled charge-coupled device (CCD) detector.

The $RC$ spectrum was obtained as $(I_S - I_r)/I_r$, where $I_S$ represents the reflection spectrum of interest, and $I_r$ represents a featureless reflection spectrum obtained from the sample upon high hole doping or from the adjacent hBN substrate. During measurements, the bilayer was grounded, and gate voltages were applied using two sourcemeters (Keithley 2400).

The PL and Raman measurements were conducted with the same back-scattering setup. A continuous-wave laser with wavelength of 532 nm was employed as the excitation light. A series of Bragg filters was used to spectrally clean the laser, and block the laser in Raman and PL spectrum. To avoid laser-induced heating, the excitation power was limited to 10 μW.

## 4. Analysis of *RC* spectra

To extract the spectral weight and resonance energy of moiré intralayer excitons, a smooth baseline—determined by a polynomial fit to the featureless regions—was first subtracted from the raw $RC$ spectra, as shown in Fig. S3(a). The excitonic resonances were then fitted using a phenomenological Lorentzian model [52] modified with a phase factor $\alpha$:

$$\frac{A}{(E-\hbar\omega_X)^2+\gamma^2/4}\left[\frac{\gamma}{2}\cos\alpha - (E-\hbar\omega_X)\sin\alpha\right],$$

where $A$, $\hbar\omega_X$, and $\gamma$ denote the spectral weight, resonance energy, and linewidth, respectively, and $E$ denotes the photon energy. $\alpha$ accounts for the asymmetric lineshape arising from optical interference effects within the van der Waals heterostructure (i.e., $\alpha = 0$ recovers a pure Lorentzian profile). As shown in Fig. S3(b), excellent fits were obtained by treating $\alpha$ as a free parameter for the lowest-energy intralayer exciton ($X_I$, as it has obvious asymmetric lineshape at low pressure) and fixing $\alpha$=0 for the higher-energy intralayer exciton resonances.

## 5. Continuum model simulation of moiré intralayer excitons

*The Hamiltonian of moiré intralayer exciton*: To elucidate the pressure dependence of moiré intralayer excitons (Fig. 2), we adopted the continuum model from ref. [25]. For monolayer $WSe_2$, the lowest-energy Hamiltonian of the center-of-mass motion of A exciton is given by,

$$H_0 = \left(E_0 + \frac{\hbar^2 \boldsymbol{Q}^2}{2M}\right)\tau_0 + J|\boldsymbol{Q}|\tau_0 + J|\boldsymbol{Q}|\left[cos(2\phi_{\boldsymbol{Q}})\tau_x + sin(2\phi_{\boldsymbol{Q}})\tau_y\right],$$

where $\boldsymbol{Q}$ is exciton momentum, $M$($\approx$1 $m_0$, $m_0$ is the free electron mass) [40] is the exciton effective mass, $E_0$ is the zero-momentum exciton energy, $J$(0.04 eV·nm) [25] accounts for the exchange interaction, $\tau_0$ and $\tau_{x,y}$ respectively are the identity and Pauli matrices in the valley pseudospin space, $\phi_{\boldsymbol{Q}}$ is the orientation angle of the 2D vector $\boldsymbol{Q}$. We use constant values of $J$ and $M$ for all pressures, which will be justified at the end of this section. For a twisted bilayer, this monolayer Hamiltonian is modulated by a periodic moiré potential $\Delta(\boldsymbol{r})$:

$$H = H_0 + \Delta(\boldsymbol{r}) = H_0 + \sum_{j=1}^{6} V e^{i\Psi_j}\exp(i\boldsymbol{b}_j\cdot\boldsymbol{r}),$$

where $\boldsymbol{b}_j$ are the moiré reciprocal lattice vectors (corresponding to a triangular moiré lattice constant of 7.7 nm) [2], and $\Delta(\boldsymbol{r})$ is approximated by the first harmonic expansion with the amplitude $V$ and the phase $\Psi_j \equiv (-1)^j\Psi$.

*Calculation of theoretical optical spectra*: To determine the optical conductivity of moiré intralayer exciton, we calculated the exciton moiré band structure by diagonalizing $H$ for given $(V,\ \Psi)$ using a plane-wave expansion basis [Fig. S8(a),(c)]. Because the photon momentum is negligible, the real part of the optical conductivity, $\mathrm{Re}\,\sigma(\omega)$, is governed by the $\gamma$ -point eigenstates $|\chi_n\rangle$ [Fig. S8(b),(d)]. $\mathrm{Re}\,\sigma(\omega)$ was calculated by summing the contributions from all $|\chi_n\rangle$, dictated by their projection onto the zero-momentum states in the absence of a moiré potential, $\left|K_\alpha^{(0)}\right\rangle$ ($\alpha = \pm$, denotes the valley index):

$$\mathrm{Re}\,\sigma(\omega) \propto \sum_n \left|\sum_{\alpha=\pm}\left\langle\chi_n\middle|K_\alpha^{(0)}\right\rangle\right|^2 \Gamma(\omega-\omega_n),$$

where $\Gamma$ is a Lorentzian broadening function, and $\omega_n$ is the angular frequency of $|\chi_n\rangle$. For simplicity, a uniform linewidth of 2 meV was assigned to all moiré intralayer excitons.

*Simulation of experimental spectra*: The experimental spectra were simulated by systematically varying the parameters $V$ and $\Psi$ in the continuum model to match four spectral features: the energy separation $E_{\mathrm{III}} - E_{\mathrm{I}}$ and three spectral weight ratios $A_i/A_{\mathrm{I}}$ ($i$ = II, III, IV), as shown in Fig. S9. We first identified the feasible parameter space for each spectral feature, defined as regions where theoretical values deviate from experimental value by less than 30% [Fig. S9(a)-(c)], or if a resonance $X_i$ was unresolved experimentally [e.g., $X_{IV}$ in Fig. S9(d)], its corresponding ratio $A_i/A_{\mathrm{I}}$ was constrained to be below 10%. We then determined the optimal parameter regime as the intersection of these individual spaces [Fig. S9(e)]. The representative center of this overlapping region was chosen as the most representative pair $(V^*, \Psi^*)$, determined as shown in Fig. S9(g): the $V^*$ was the value minimizing $\sum_{j=1}^{n}\left(V^* - V_j\right)^2$ across the optimal domain, and $\Psi^*$ was calculated as the average phase at fixed $V^*$. To be noted, absolute resonance energies are not used as fitting constraints, because they are subjected to uniform compression effects independent of moiré potential [as demonstrated by the misaligned bilayer, Fig. 2(a)].

*Calculation of Chern number*: Following ref. [25], the Berry curvature $\mathcal{F}_n(Q)$ of the *n*-th exciton moiré band is given by the curl of the Berry connection $A_n(\boldsymbol{Q})$:

$$\mathcal{F}_n(\boldsymbol{Q}) = \hat{z}\cdot\nabla_{\boldsymbol{Q}} \times A_n(\boldsymbol{Q}) = \hat{z}\cdot\nabla_{\boldsymbol{Q}} \times \left[i\langle\chi_n(\boldsymbol{Q})|\nabla_{\boldsymbol{Q}}|\chi_n(\boldsymbol{Q})\rangle\right]$$

where $|\chi_n(\boldsymbol{Q})\rangle$ is the state of the *n*-th band with momentum $\boldsymbol{Q}$. The Chern number $C_n$, is obtained by integrating the Berry curvature over the first moiré Brillouin zone (MBZ):

$$C_n = \int_{\mathrm{MBZ}} \frac{d^2\boldsymbol{Q}}{2\pi}\mathcal{F}_n(\boldsymbol{Q})$$

To yield meaningful Chern number, a finite valley Zeeman field (e.g., $h_z$=1.5 meV), has to be introduced to the Hamiltonian to break time-reversal symmetry and lift the valley degeneracy at the $\gamma$ point. See the resulting Chern-number phase diagram in Fig. S10.

*Spectral response of pressure modulations of $J$ and $M$*: We have used constant values of $J$ and $M$ in the simulation for all pressures; this can be justified by evaluating their pressure-induced variations and the corresponding spectral response. Following Ref. [53], $J \propto |\psi_{eh}(0)|^2/\epsilon$, where $|\psi_{eh}(0)|^2$ is the probability of the spatial overlap between the electron and the hole, and $\epsilon$ is the effective dielectric constant of hBN. According to a two-dimensional hydrogenic mode [54], $|\psi_{eh}(0)|^2 \propto \epsilon^{-2}$, and hence $J \propto \epsilon^{-3}$. Theoretical calculations [55] show that the hBN dielectric constant increases by ~6% as pressure increases to 3.6 GPa, which translates into a 16% reduction of $J$. For $M$, the calculated conduction- and valence-band effective masses at the $K$ point exhibit no discernible change up to 3 GPa [55], indicating that $M$ remains essentially constant over the pressure range explored here.

Fig. S20(a) and (b) illustrate the evolution of calculated optical spectra upon varying $J$ and $M$, respectively, with all remaining parameters held at their 0.5 GPa values ($V$=29 meV and $\Psi$=23°). The corresponding energy separations and relative spectral weights are extracted in Fig. S20(c) and (d). To perform a thorough sensitivity analysis, both parameters are varied over ranges considerably wider than the estimates above: $J$ by $\pm$20% around 40 meV·nm; $M$ by $\pm$10% around 1 $m_0$. Across these ranges, moiré excitons exhibit only weak spectral variations, which are insufficient to account for the pronounced pressure-driven evolution observed experimentally (Fig. 2 of the main text).

## 6. Pressure dependence of intralayer excitons in the $WS_2$ spectral range

Fig. S4(a) shows the pressure dependence of *RC* spectra from 1.9 eV to 2.3 eV, covering the spectral range of the $WS_2$ intralayer exciton. At ambient pressure, three prominent excitonic resonances are observed, denoted as $X_1$, $X_2$, and $X_3$. As the pressure increases to 3.6 GPa, all three resonances redshift; concurrently, the $X_1 - X_2$ energy separation decreases, and the relative spectral weights of $X_2$ and $X_3$ with respect to $X_1$ increase [Fig. S4 (b)].

A quantitative continuum-model analysis of these resonances is limited by a substantial spectral overlap with the $WSe_2$ B exciton ($X_B^{Se}$). Using the misaligned bilayer as a reference (Fig. S21), we estimate the energy of $X_B^{Se}$ in the 60°-twisted bilayer to be ~2.10 eV, which coincides with $X_2$. This overlap prevents extraction of the intrinsic spectral profile of the $WS_2$ moiré excitons from the measured spectra. A similar limitation has also hindered a detailed analysis of the $WS_2$ spectral range in a previous study of $WSe_2$/$WS_2$ moiré bilayers [40].

## 7. Discussion on the $X_{IV}$ resonance

The $X_{IV}$ resonance observed in *RC* spectra at elevated pressures (Fig. 2) is attributed to several higher-energy moiré bands of intralayer exciton, as shown Fig. S8. At low pressure, the states at $\gamma$-point states associated with $X_{IV}$ possess negligible component of $\left|K_\alpha^{(0)}\right\rangle$, rendering them optically dark. However, they gain significant zero-momentum weight at higher pressures, becoming optically active. Furthermore, the intralayer nature of $X_{IV}$ is corroborated by its negligible energy shift under an applied out-of-plane electric field (Fig. S22), in contrast to the behavior of the spatially hybridized exciton $X_h$.

## 8. Image-charge model for doping-induced intralayer exciton energy shifts

Upon electron doping (Fig. 3 of the main text), $WS_2$ becomes metallic and thereby imposes Coulomb screening on the intralayer excitons of $WSe_2$. We model the screening effects using the image-charge method, initially developed for molecules adsorbed on metal surfaces [45] and later applied to screened excitons in 2D semiconductors [56,57]. As illustrated in Fig. S12(a), the $WSe_2$ exciton is represented as an electron and a hole separated by an in-plane distance *r*; both induce image charges in the adjacent metallic $WS_2$ layer, located at a perpendicular distance *d*. The screening-induced shifts are,

$$\Delta E_g = -\frac{C}{2d}, \qquad \Delta E_{\mathrm{B}} = -\frac{C}{\sqrt{r^2+4d^2}} .$$

Here, $C = \frac{\eta e^2}{4\pi\epsilon_0\epsilon_{\mathrm{eff}}}$, where $e$ is the elementary electron charge, $\epsilon_0$ is the vacuum permittivity, $\epsilon_{\mathrm{eff}}$ is the effective surrounding dielectric constant, and $\eta$ (varying from 0 to 1) characterizes the screening strength of $WS_2$. The corresponding net shift of the exciton energy is given by, $\Delta E_{\mathrm{X}} = -C\left[\frac{1}{2d} - \frac{1}{\sqrt{r^2+4d^2}}\right]$, which is always negative for all *r*. Therefore, this framework naturally yields a net redshift. It is worth emphasizing that this model is highly simplified and cannot account for the screening-induced blue shift of $X_{\mathrm{II}}$. For instance, this model does not include the spatially non-uniform electron distribution in $WS_2$ arising from the moiré superlattice, which can modify the exciton moiré potential of $WSe_2$ and consequently reshape the spectrum.

We further apply this model to understand the physics picture behind the pressure evolution of the doping-induced shift of $X_{\mathrm{III}}$ [Fig. 3(f)]. In general, loosely bound excitons with large electron-hole spatial separations—such as Rydberg and charge-transfer excitons—exhibit larger screening-induced shifts [45,58]. This trend is captured by the image-charge model: the magnitude of the exciton redshift increases with *r* (for fixed *d*) [Fig. S12 (b)]. Consequently, the decrease of the screening-induced redshift of $X_{\mathrm{III}}$ upon pressure [Fig. 3(f)], suggests a reduction of *r*, which implies that $X_{\mathrm{III}}$ evolves from a charge-transfer exciton to a more tightly bound exciton [Fig. 3 (g)].

## 9. Pressure dependence of phonon modes in the 60°-twisted bilayer

To confirm the structural stability and reversibility, we performed the Raman measurement throughout the entire pressure cycle. Fig. S18(a) shows the pressure dependence of phonon modes of $WSe_2$, denoted as $E_{2g}$, $A_{1g}$, and 2LA. All phonon modes exhibit blue-shifts linearly with the pressure [Fig. S18(b)], owing to strengthening of atomic bonds; importantly, all mode energies fully recover upon decompression, signifying negligible pressure-induced sample damage. To be noted, the linewidth of $A_{1g}$ mode becomes slightly broadened above 3 GPa [Fig. S18(c)], indicating pressure-induced inhomogeneity and non-ideal hydrostaticity, which should be associated with the solidification of the PTM at cryogenic temperature.

We further investigated the pressure dependence of the low-frequency Raman mode, providing complementary experimental support for the pressure-induced enhancement of interlayer coupling. As shown in Fig. S23, two low-frequency modes are observed, denoted as $P_1$ and $P_2$. Both modes are $A_1$-symmetric: they appear only in the parallel-linear (XX) polarization channel, and are absent in the perpendicular-linear polarization (XY) channel. As pressure increases, $P_1$ exhibits nearly unchanged energy around 22 $cm^{-1}$, whereas $P_2$ exhibits a blueshift, reaching about 37 $cm^{-1}$ at 5.7 GPa [Fig. S23(b)].

We attribute $P_2$ to the interlayer breathing mode. Its pressure-induced hardening is consistent with that the pressure reduces the interlayer distance and thereby strengthens the interlayer coupling [59]. The nearly pressure-independent energy of $P_1$ suggests that it might not be an interlayer breathing mode. We attribute $P_1$ to a moiré phonon—a momentum-folded acoustic mode associated with the moiré reciprocal lattice vector [60].

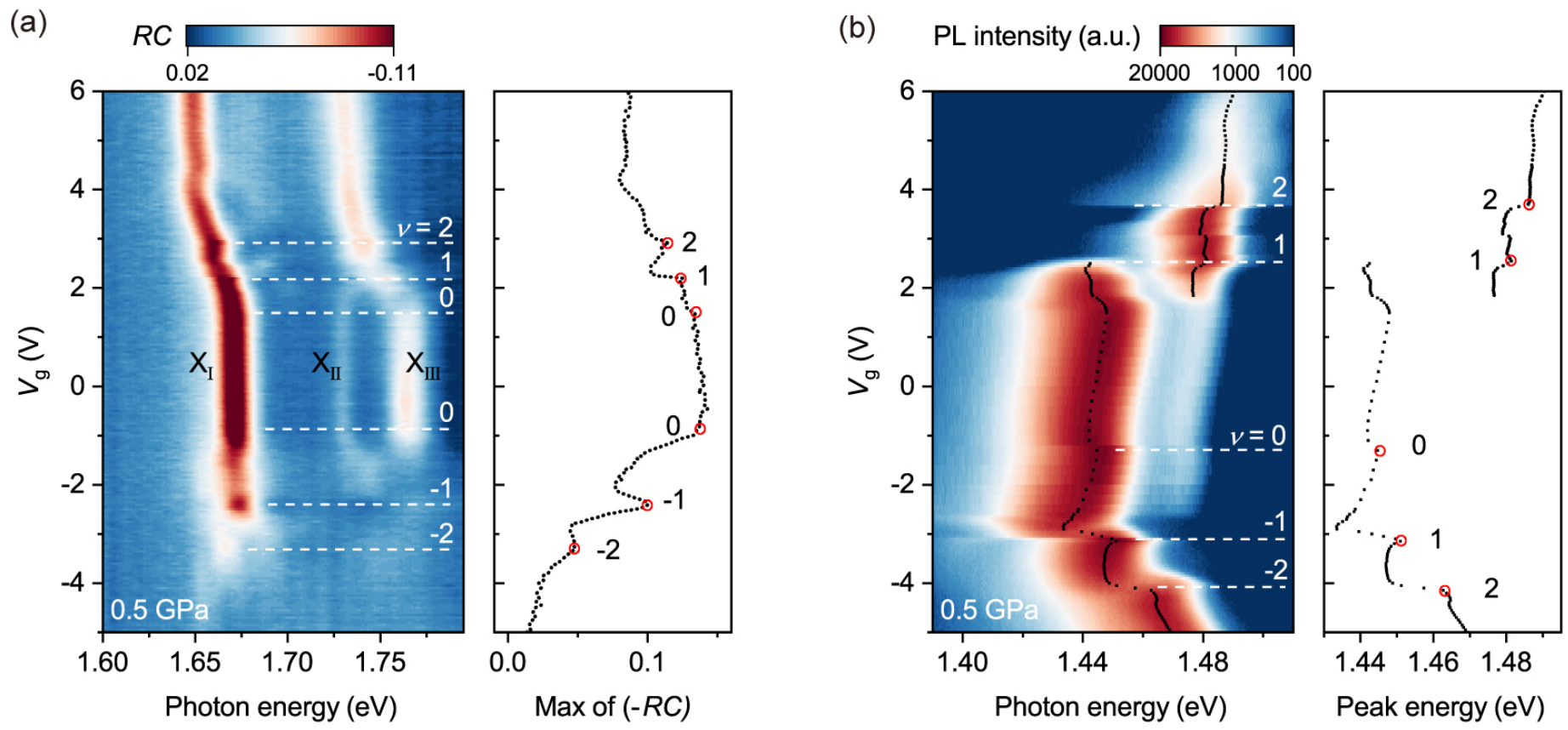

Fig. S1. Filling assignment based on the $RC$ and PL spectra. (a) Doping-dependent $RC$ spectra (left) and extracted maximum of $-RC$ around the $X_I$ resonance (right). Fillings are assigned according to the enhanced $RC$ magnitude, driven by reduced screening upon the formation of insulating states. (b) Doping-dependent PL spectra (left) and extracted IX energy (right). Fillings are assigned according to the abrupt energy shift. This is also consistent with the formation of insulating states, which could strongly modulate the physical properties of the system (e.g., dielectric condition and the Fermi sea) and hence influence the PL energy. The filling assignments are consistent with previous studies [17,27,41]. The gate-voltage offset for identical filling between PL and $RC$ measurements is attributed to the laser excitation in PL measurement, i.e., the laser likely modifies the electrode contact conditions.

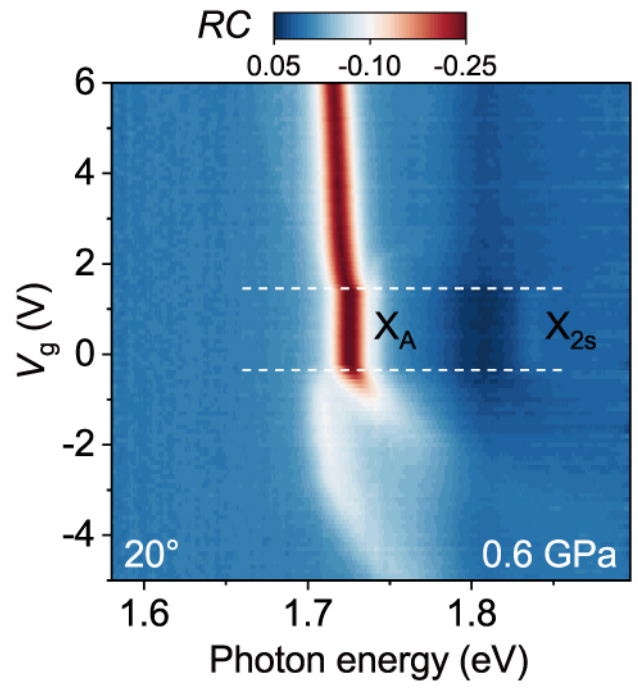

Fig. S2. *RC* spectra of the 20°-twisted device at 0.6 GPa. $X_A$ and $X_{2s}$ denote A exciton of $WSe_2$ without moiré effects and its Rydberg state, respectively. The dashed lines denote doping onset.

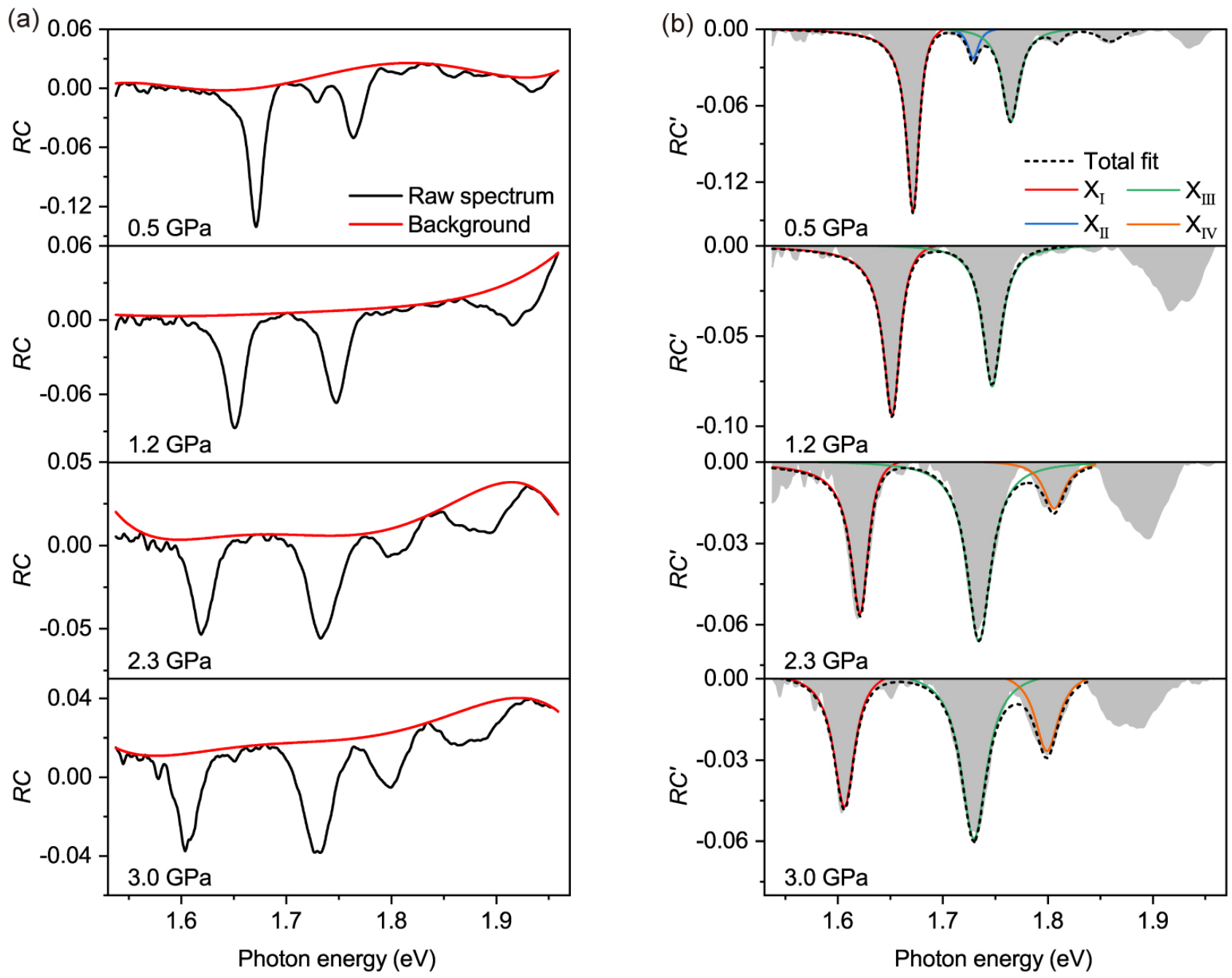

Fig. S3. Background subtraction and Lorentzian fitting of $RC$ spectra. (a) Raw $RC$ spectra and the overlaid smoothly varying backgrounds (red curves) at representative pressures. (b) Background-subtracted $RC$ spectra, denoted as $RC'$, fitted with the phenomenological Lorentzian model.

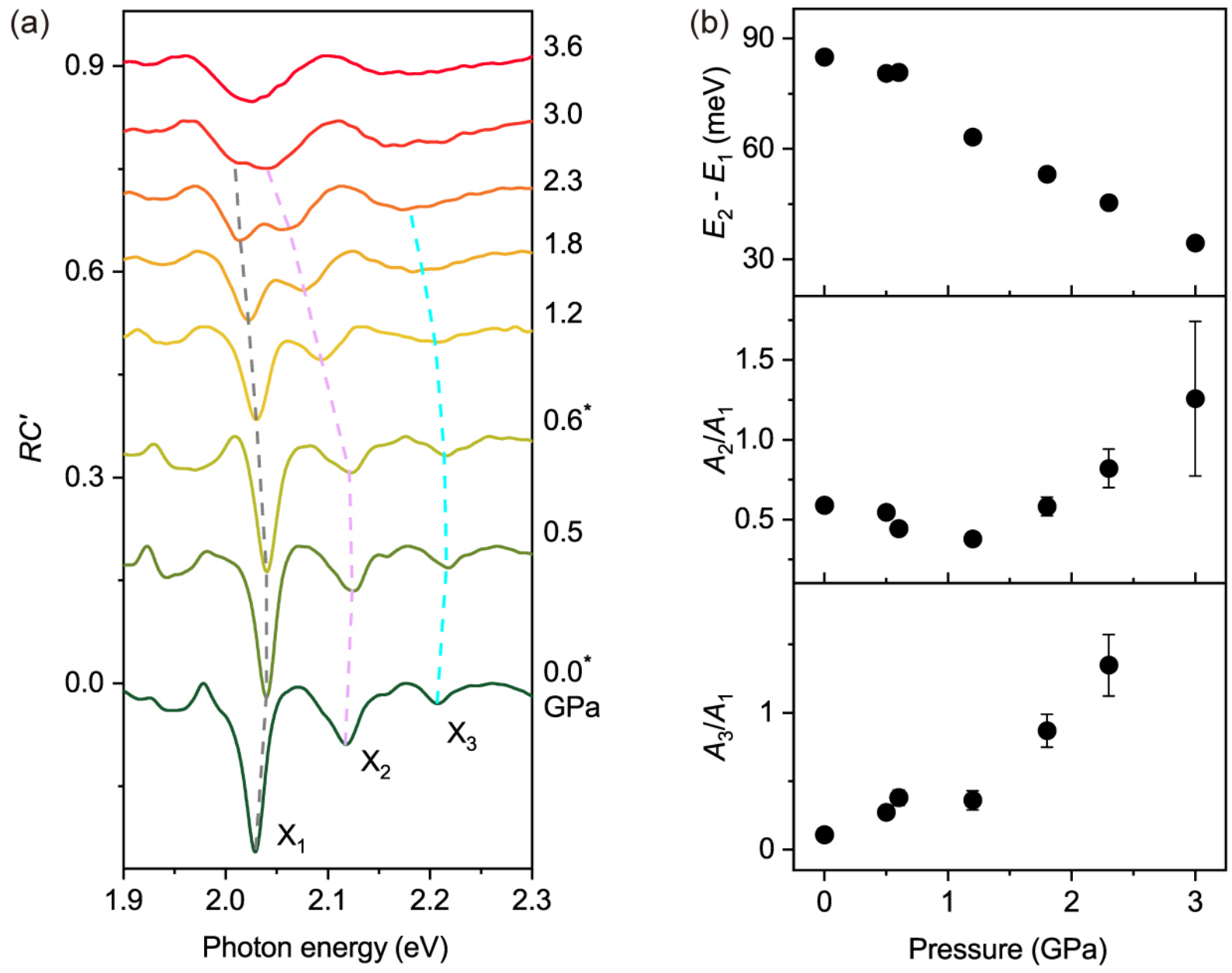

Fig. S4. Pressure-dependent exciton resonances in the $WS_2$ A-exciton energy range. (a) Background-subtracted reflectance contrast spectra in the $WS_2$ A-exciton energy range at charge neutrality as a function of pressure. Spectra are vertically offset for clarity. $X_1$, $X_2$, and $X_3$ denote three exciton resonances, and their pressure evolutions are indicated by dashed lines. (b) Pressure dependence of the spectral features of these resonances. $E_i$ and $A_i$ denote the energy and spectral weight of $X_i$ (i=1,2,3), respectively.

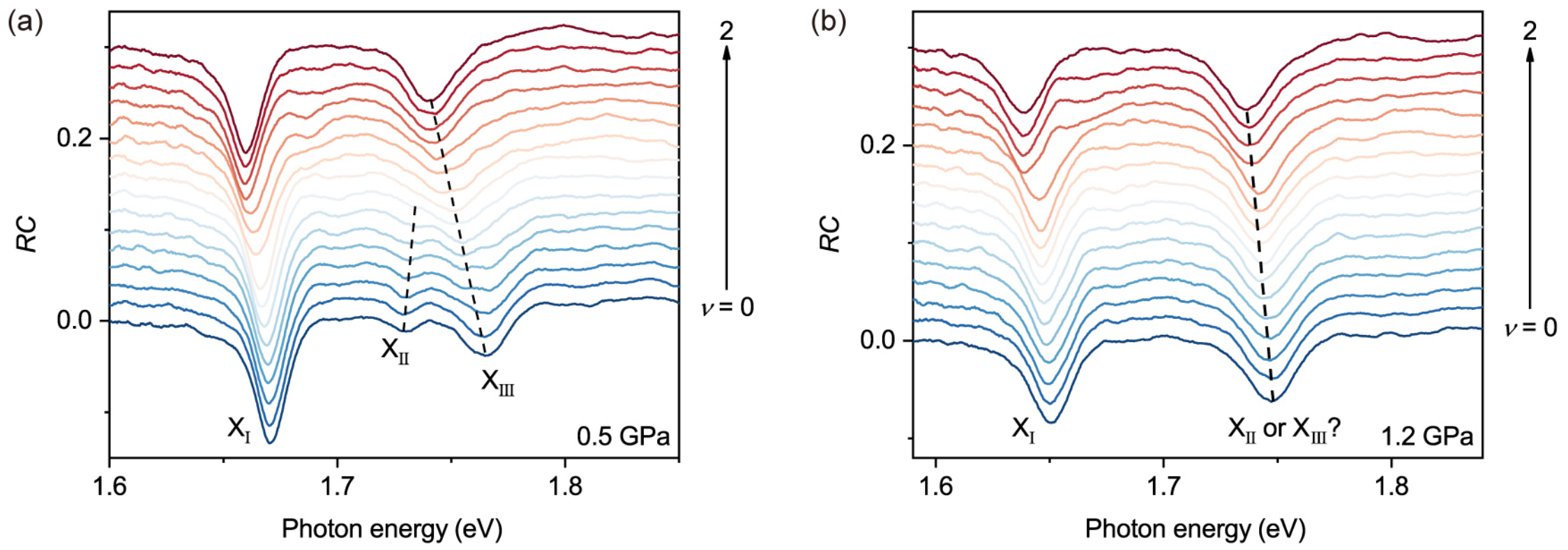

Fig. S5. Doping-dependent reflectance contrast spectra at 0.5 GPa (a) and 1.2 GPa (b). Spectra are vertically offset for clarity. Dashed lines guide the eye to the evolution of the exciton resonances. At 0.5 GPa, where all three moiré excitons are clearly identified, $X_{II}$ blueshifts with increasing electron density, whereas $X_{III}$ redshifts. At 1.2 GPa, the second-lowest-energy resonance redshifts with electron density, matching the doping behavior of $X_{III}$.

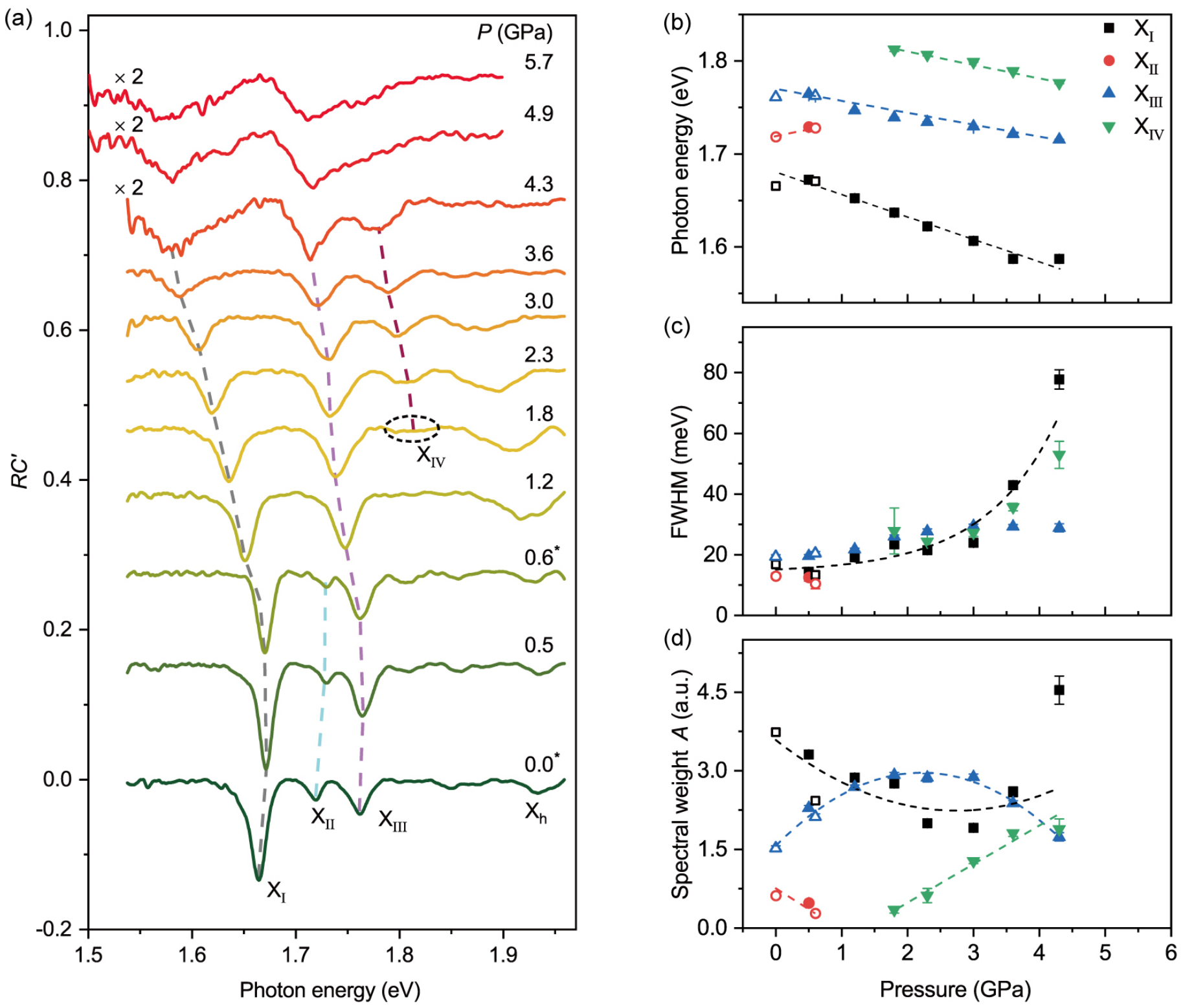

Fig. S6. Pressure evolution of moiré intralayer excitons at charge neutrality. (a) Pressure-dependent *RC* spectra at neutral doping. The spectra at high pressures are scaled by a factor of 2. All spectra are vertically displaced for clarity. (b)-(d) Pressure-dependent peak energy, full-width at half-maximum (FWHM), and spectral weight. Dashed lines are guides to the eye. Decompression data are denoted by asterisk in (a) and empty symbols in (b)-(d).

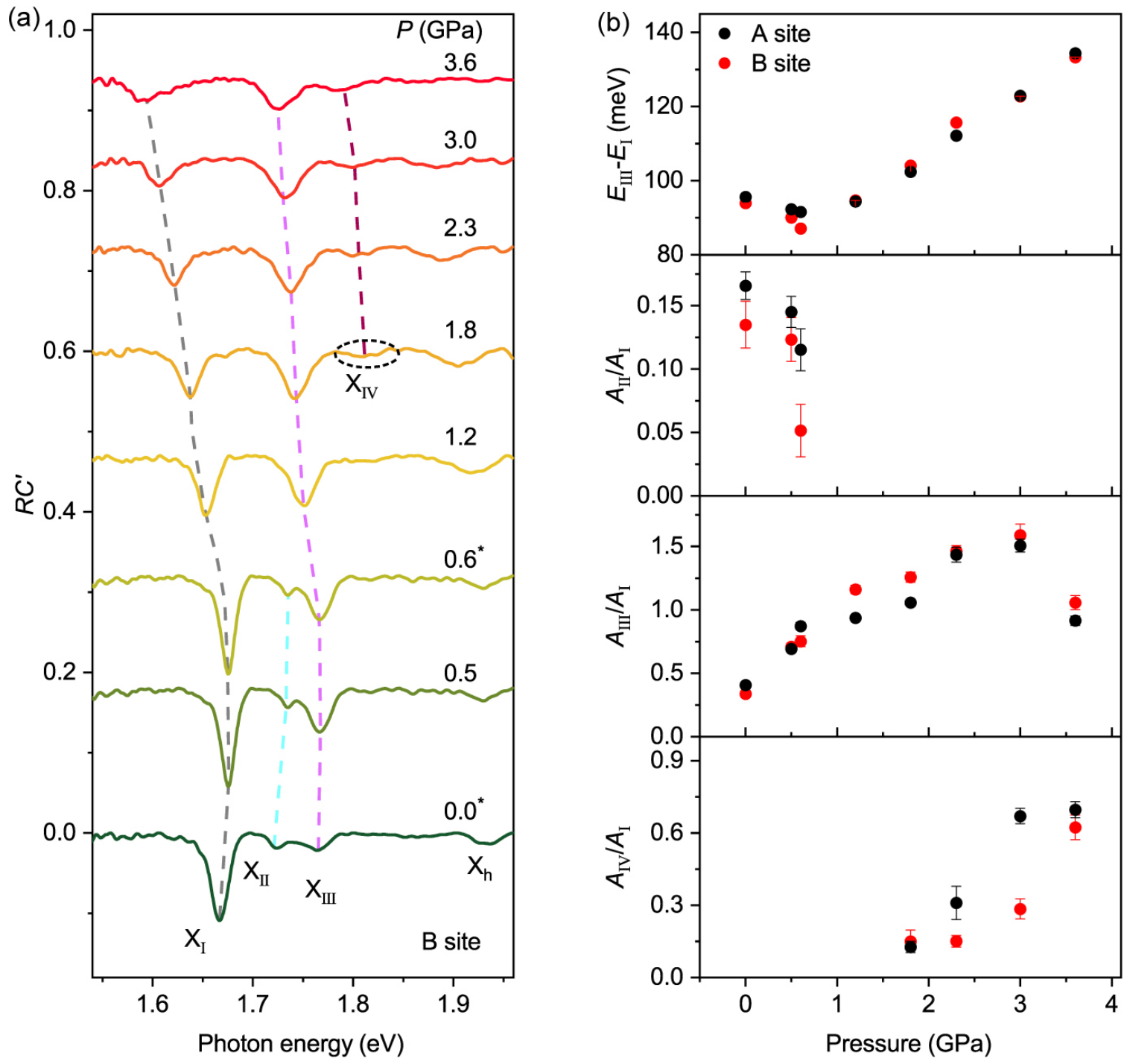

Fig. S7. Reproducibility of the pressure dependence of moiré intralayer excitons at B site of the 60°-twisted device. (a) Pressure evolution of background-subtracted $RC$ spectra at charge neutrality acquired from the B site. Spectra are vertically offset for clarity, and decompression data are denoted by asterisks. (b) Comparison between spectral features from the B site and the primary measurement location (A site; e.g., main-text Fig. 2). Both spatial locations exhibit highly consistent pressure evolution of the moiré intralayer excitons.

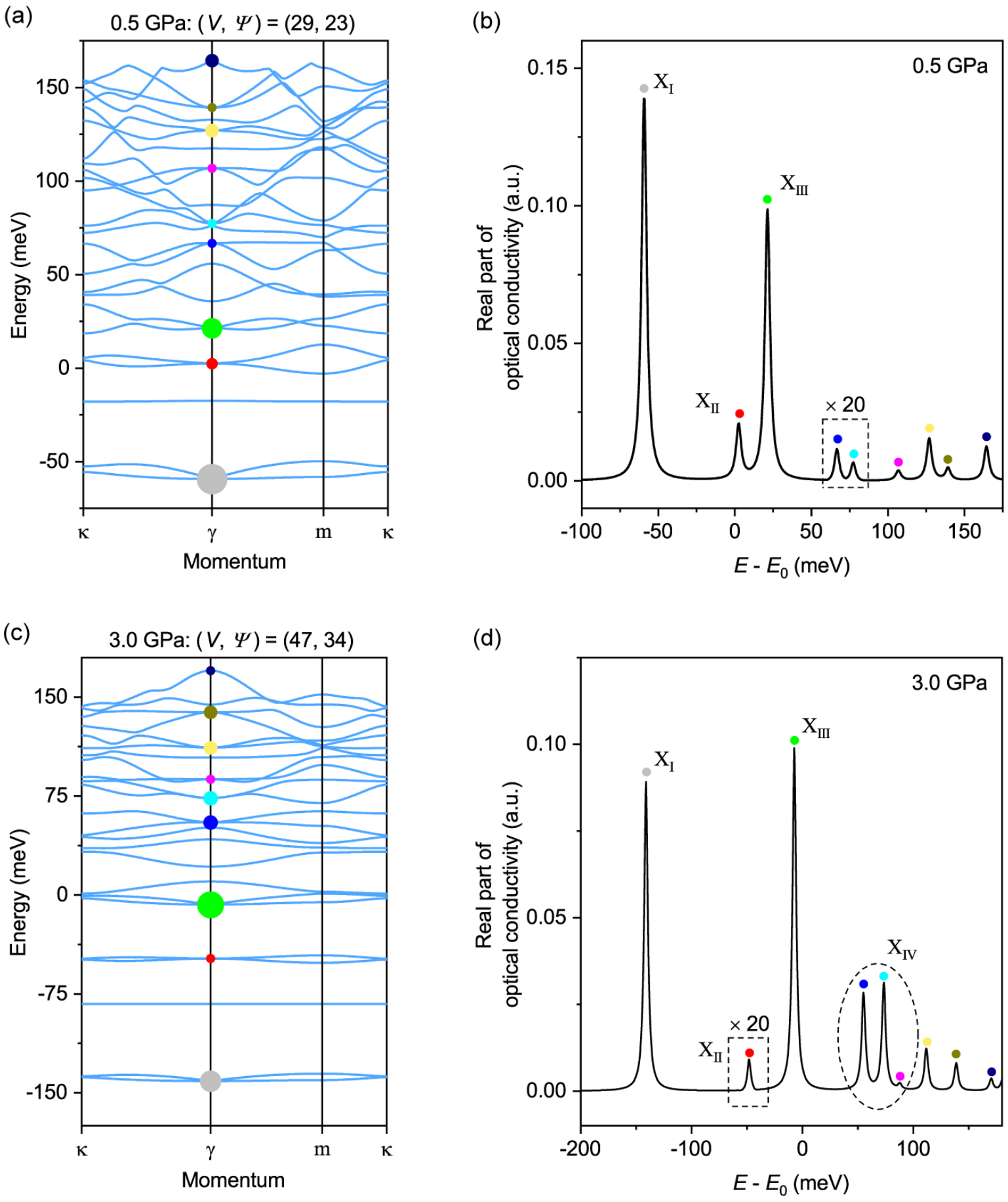

Fig. S8. Theoretical moiré exciton band structures (a,c) and optical conductivity spectra (b,d) at 0.5 and 3.0 GPa. The moiré-potential parameters $V$ and $\Psi$ (top) are extracted from the representative centers in Fig. 2(e). Colored circles (a,c) denote $\gamma$-point states that contribute to the optical conductivity, with the size indicating the magnitude of the oscillator-strength magnitude.

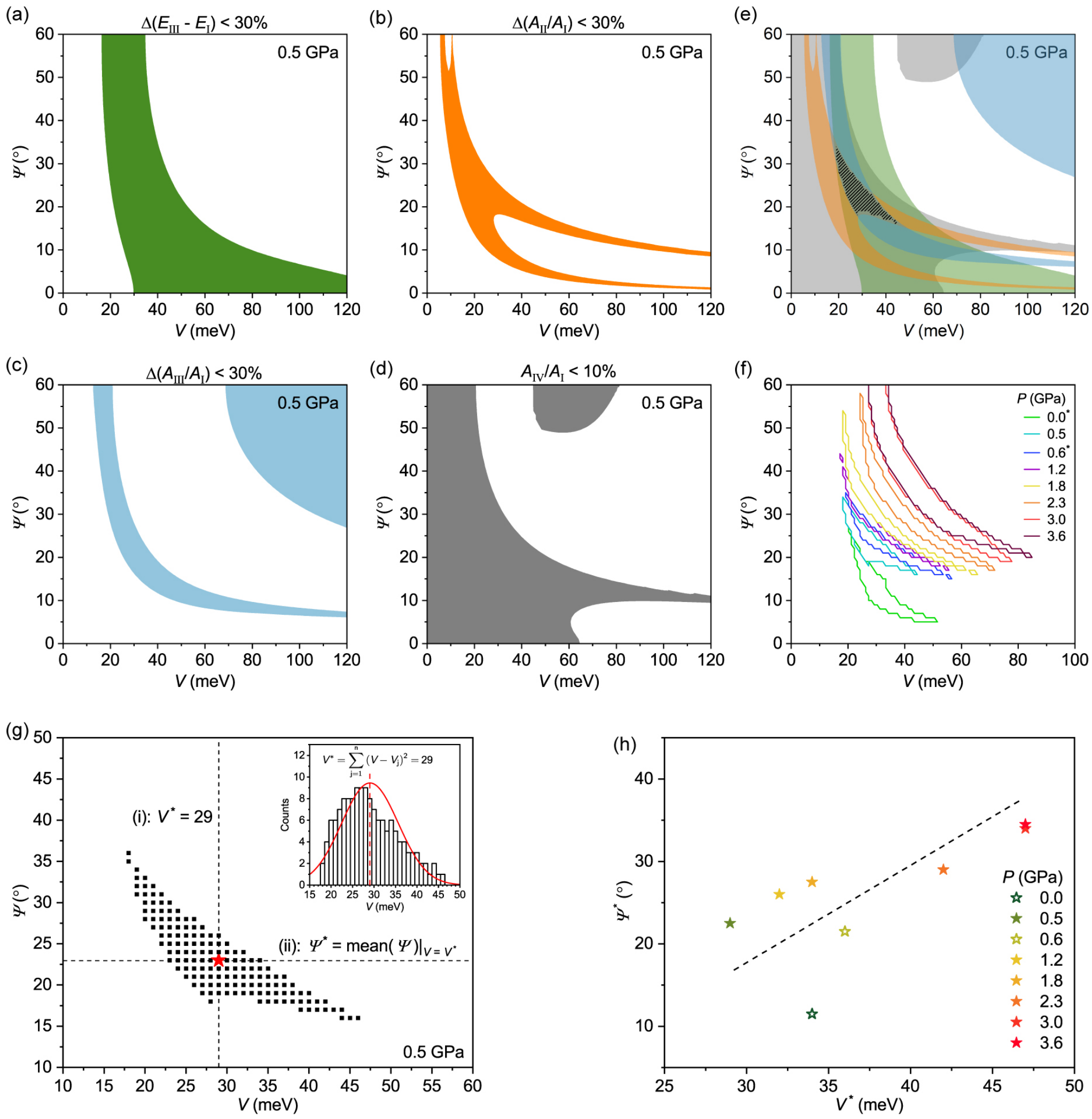

Fig. S9. Determination of the moiré-potential parameter regime of $(V, \Psi)$ and the corresponding representative center. (a)-(d) Taking 0.5 GPa as an example, Parameter sub-regions of four spectral characteristics—$E_{\rm III} - E_{\rm I}$ and $A_i/A_{\rm I}$ ($i$ = II, III, IV), constrained by experimental tolerances at a representative pressure of 0.5 GPa. For instance, the relative deviation $\Delta(E_{\rm III} - E_{\rm I}) \equiv \left| {(E_{\rm III} - E_{\rm I})_{theor.}}\big/{(E_{\rm III} - E_{\rm I})_{exp.}} - 1 \right|$ is bounded below 30%. As $X_{\rm IV}$ cannot be experimentally detected at 0.5 GPa, $A_{IV}/A_{\rm I}$ is restricted below 10%. (e) Optimal parameter regime (the hatched area), determined by the intersection of the four characteristics constraints. (f) Pressure-dependent evolution of the optimal parameter space. (g),(h) Extraction of the representative center (star) within the optimal parameter regime (g), and its pressure dependence (h). Solid and open stars correspond to compression and decompression, respectively.

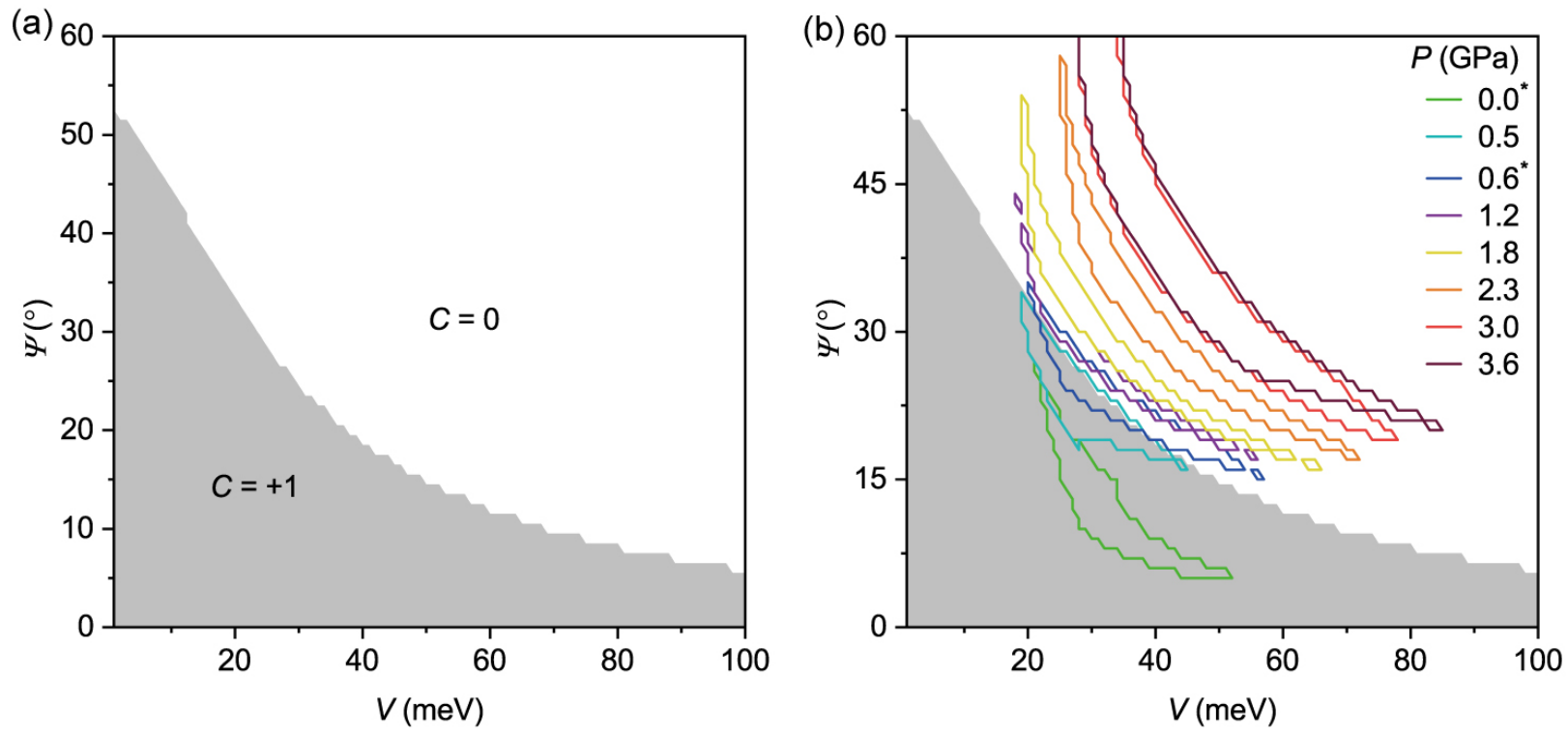

Fig. S10. The Chern-number phase diagram as a function of $V$ and $\Psi$. (a) Two distinct regimes with valley Chern number of +1 and 0. A valley Zeeman splitting ($h_z$ = 1.5 meV) is applied to the Hamiltonian of moiré intralayer exciton. (b) Pressure evolution of the optimal parameter regime (extracted from Fig. S9), superimposed onto the topological phase diagram.

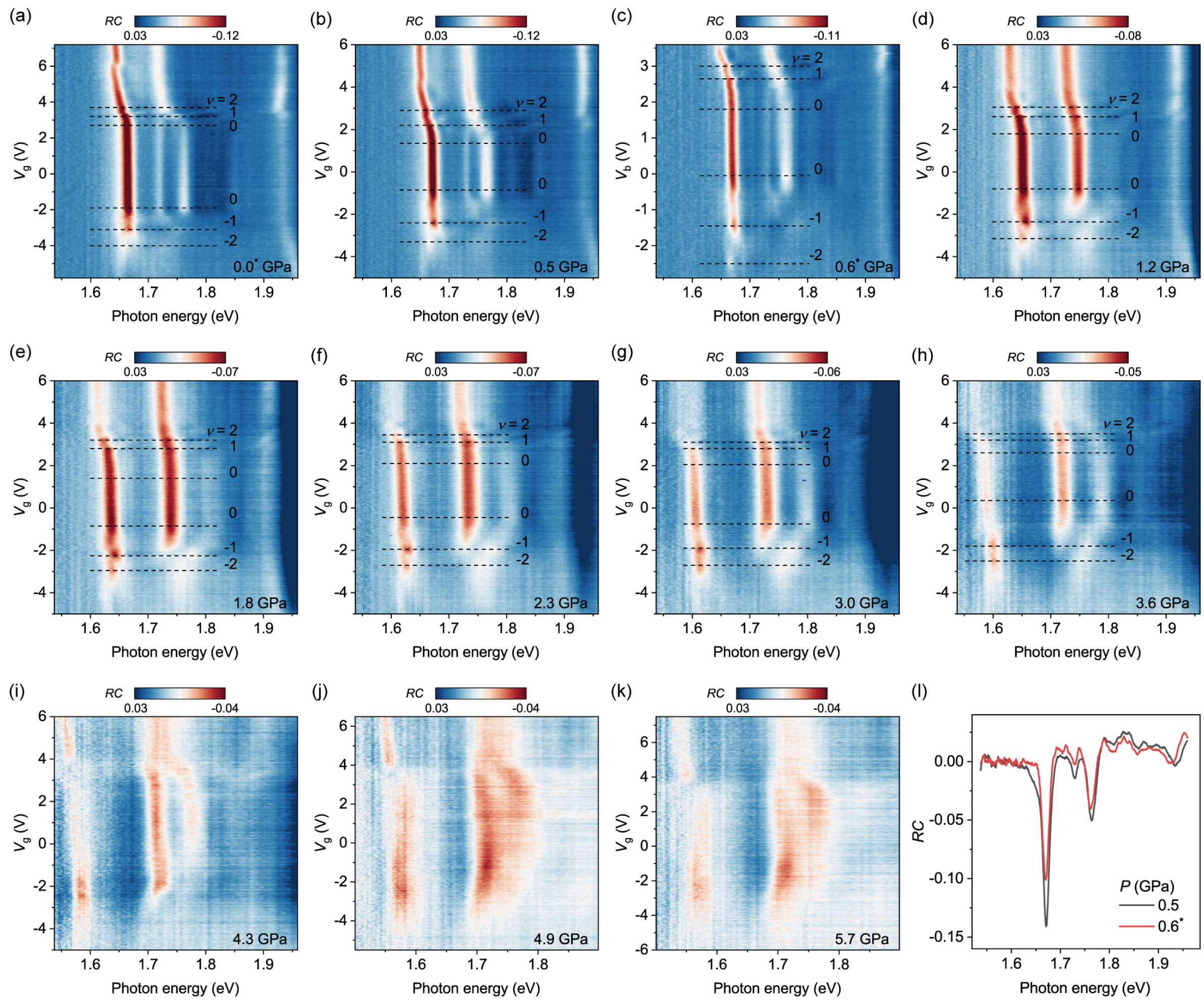

Fig. S11. Doping-dependent *RC* spectra for all the pressures and measurement reversibility. (a)-(k) Doping-dependent *RC* spectra at varying pressures. Pressure-induced variation in the gate voltage corresponding to specific fillings arise primarily from altered electrode contact conditions. (l) Comparison of charge-neutral *RC* spectra at 0.5 GPa (compression) and 0.6 GPa (decompression, denoted by an asterisk). Moiré intralayer excitons exhibit highly consistent peak energies and spectral weights during both processes, apart from a slight reduction in the $X_I$ spectral weight at 0.6 GPa.

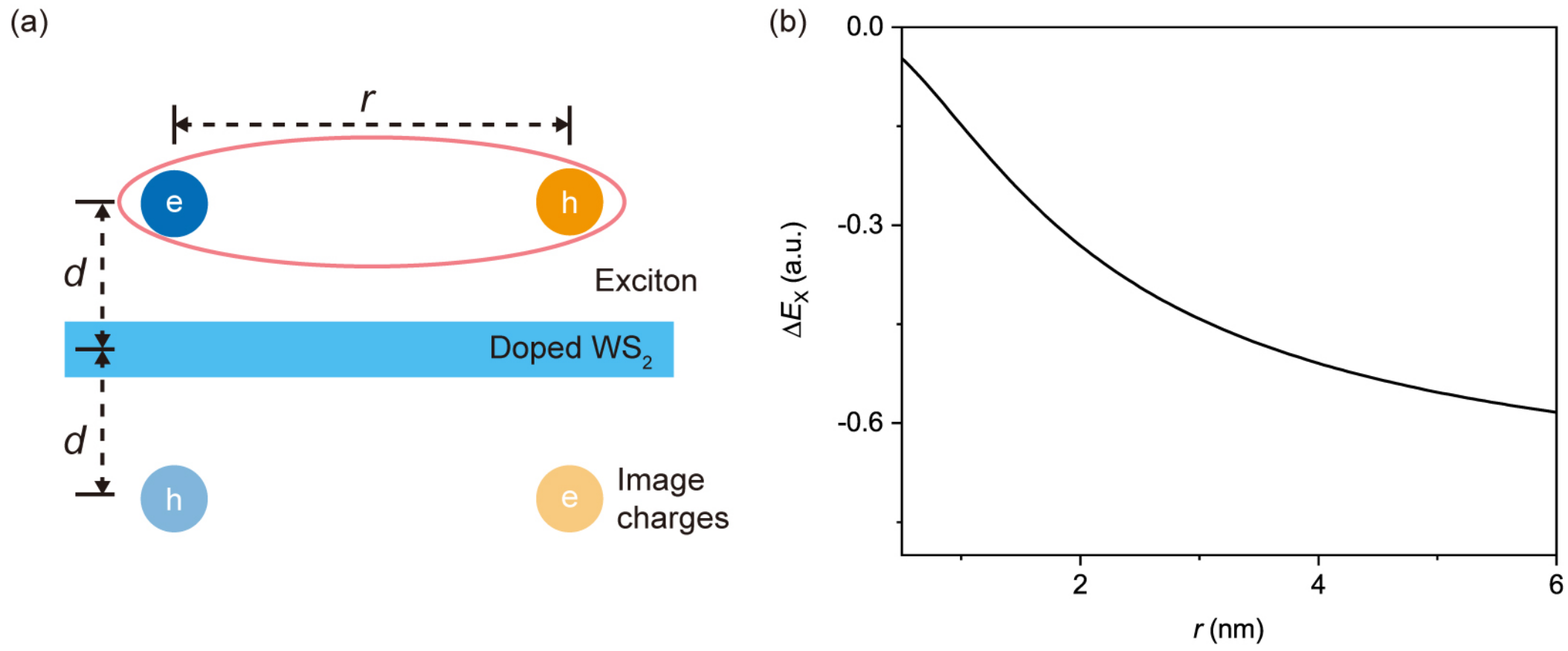

Fig. S12. The image-charge model for screening effects. (a) An electron-hole pair (i.e., exciton) screened by a doped $WS_2$ monolayer, with a perpendicular distance $d$. The electron and hole are separated by an in-plane distance $r$. The corresponding image charges reside beneath the $WS_2$. (b) Screening-induced net shift of the exciton energy, $\Delta E_X$, as a function of $r$, with $d$ fixed at 0.67 nm.

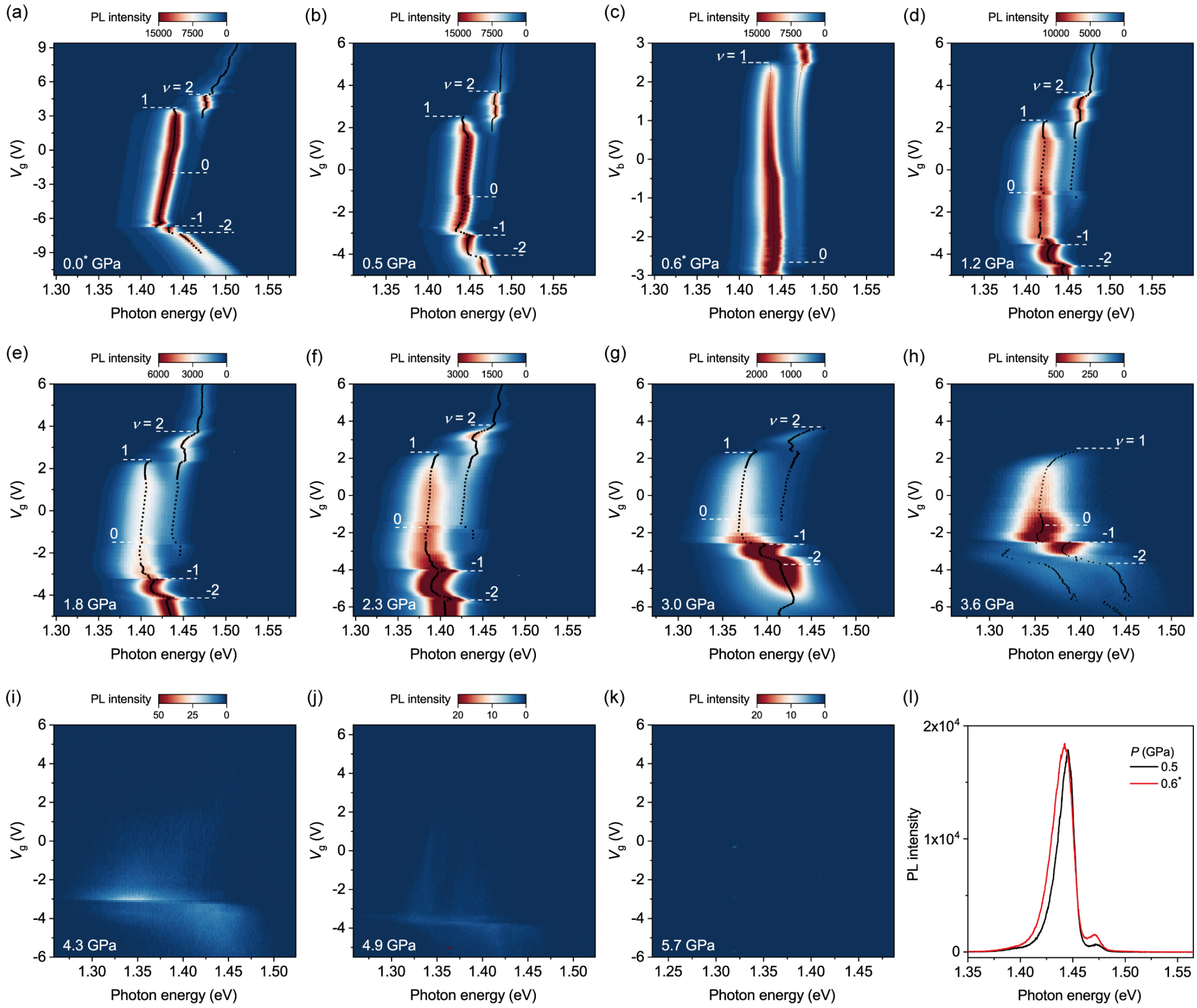

Fig. S13. Pressure evolution of doping-dependent PL spectra and measurement reversibility. (a)-(k) Evolution of the doping-dependent PL spectra at varying pressures. The gate voltage required to reach a specific filling factor shifts as a function of pressure, arising primarily from pressure-induced modifications to the electrical contact conditions. (l) Comparison of charge-neutral PL spectra at 0.5 GPa (compression) and 0.6 GPa (decompression, denoted by an asterisk). The two spectra exhibit nearly identical interlayer exciton (IX) peak energies and intensities. These two spectra are similar, in terms of the peak energy and PL intensity of the IX. The PL intensities with various pressures are normalized to the same acquisition conditions, such as laser power and exposure time.

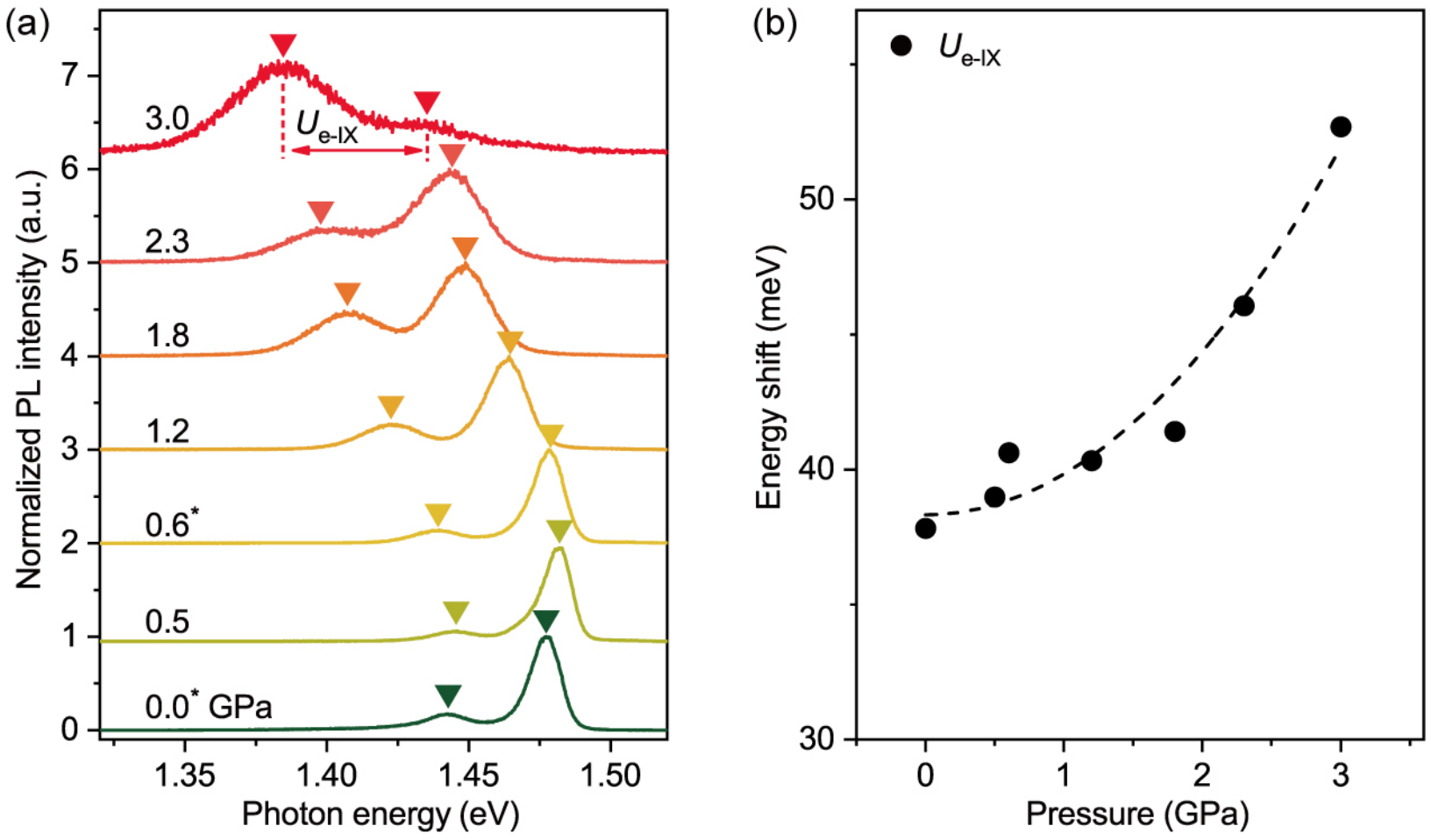

Fig. S14. Pressure-dependent determination of $U_{\text{e-IX}}$ at $\nu = 1$. (a) Evolution of the normalized PL spectra extracted at $\nu = 1$. The energy separation between the two prominent peaks, indicated by solid inverted triangles, defines $U_{\text{e-IX}}$. Spectra are vertically displaced for clarity, with decompression data marked by asterisks. (b) Pressure evolution of $U_{\text{e-IX}}$ (as shown in Fig. 4 of the main text).

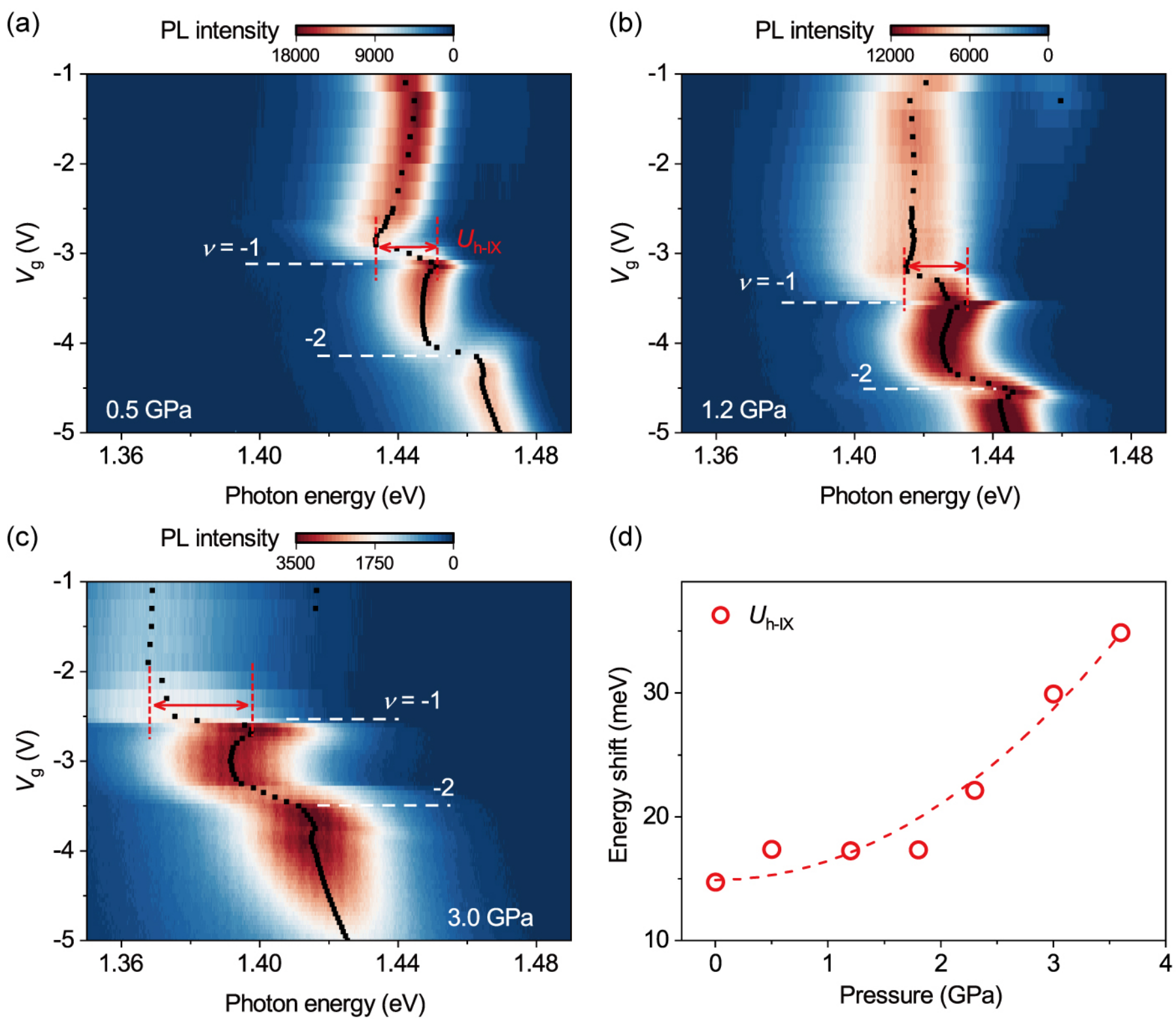

Fig. S15. Pressure-modulated interplay between doped holes and interlayer excitons. (a)-(c) Doping-dependent PL spectra at 0.5 GPa(a), 1.2 GPa(b) and 3.0 GPa(c). Dashed curves trace the IX peak energy. Red arrows denote the energy shifts around $\nu$ = -1, denoted as $U_{\text{h-IX}}$. (c) Pressure evolution of $U_{\text{h-IX}}$.

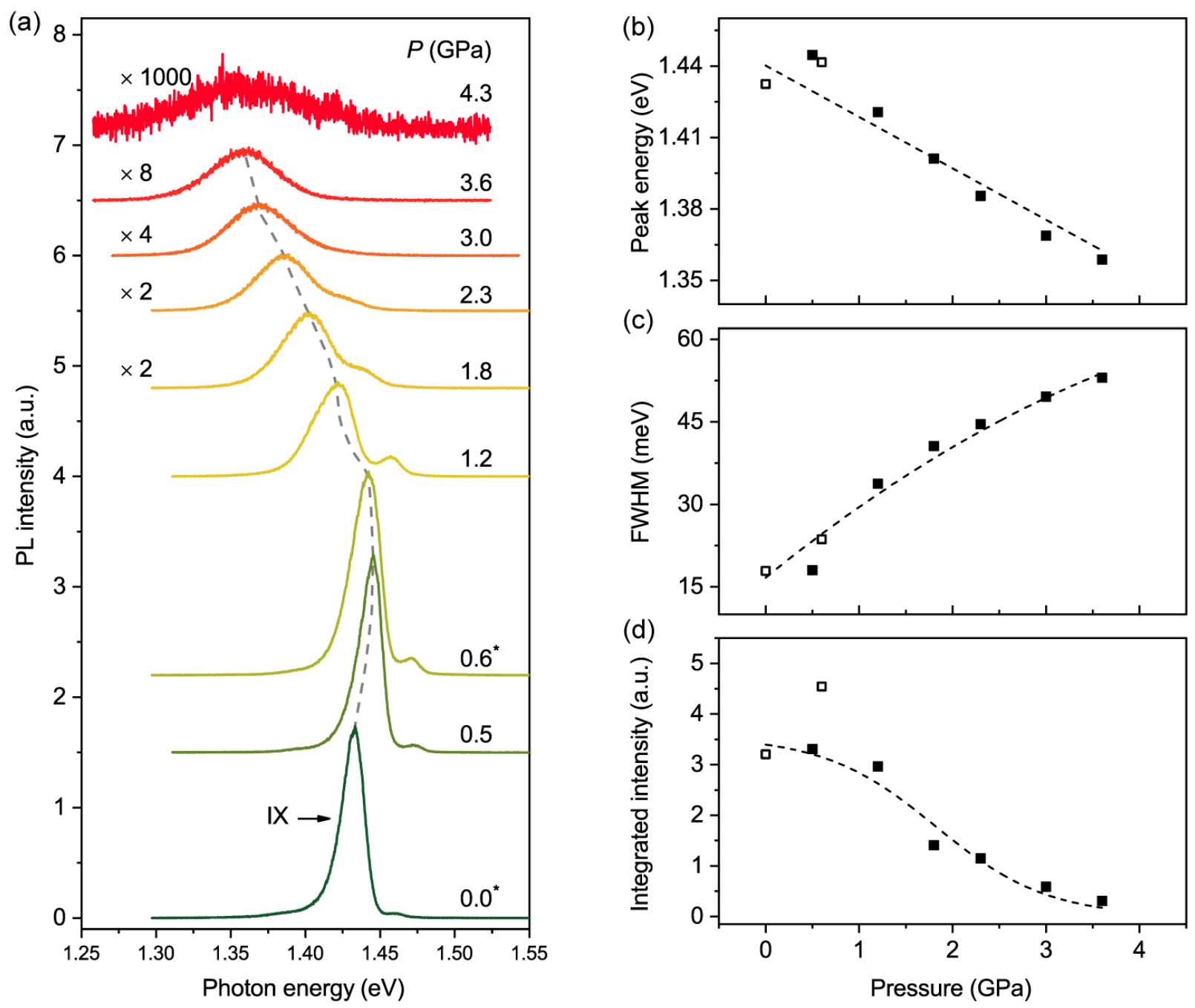

Fig. S16. Pressure-dependent PL spectra at charge neutrality. (a) Pressure evolution of the PL spectrum extracted from the charge-neutral region of Fig. S13. (b)-(d) Pressure dependence of the IX characteristics, specifically the peak energy (b), FWHM (c), and integrated intensity (d). The integrated intensities with various pressures are normalized to the same acquisition conditions, such as laser power and exposure time. Data acquired during decompression are denoted either by an asterisk (a), or open symbols (b-d). Dashed lines are guides to the eye.

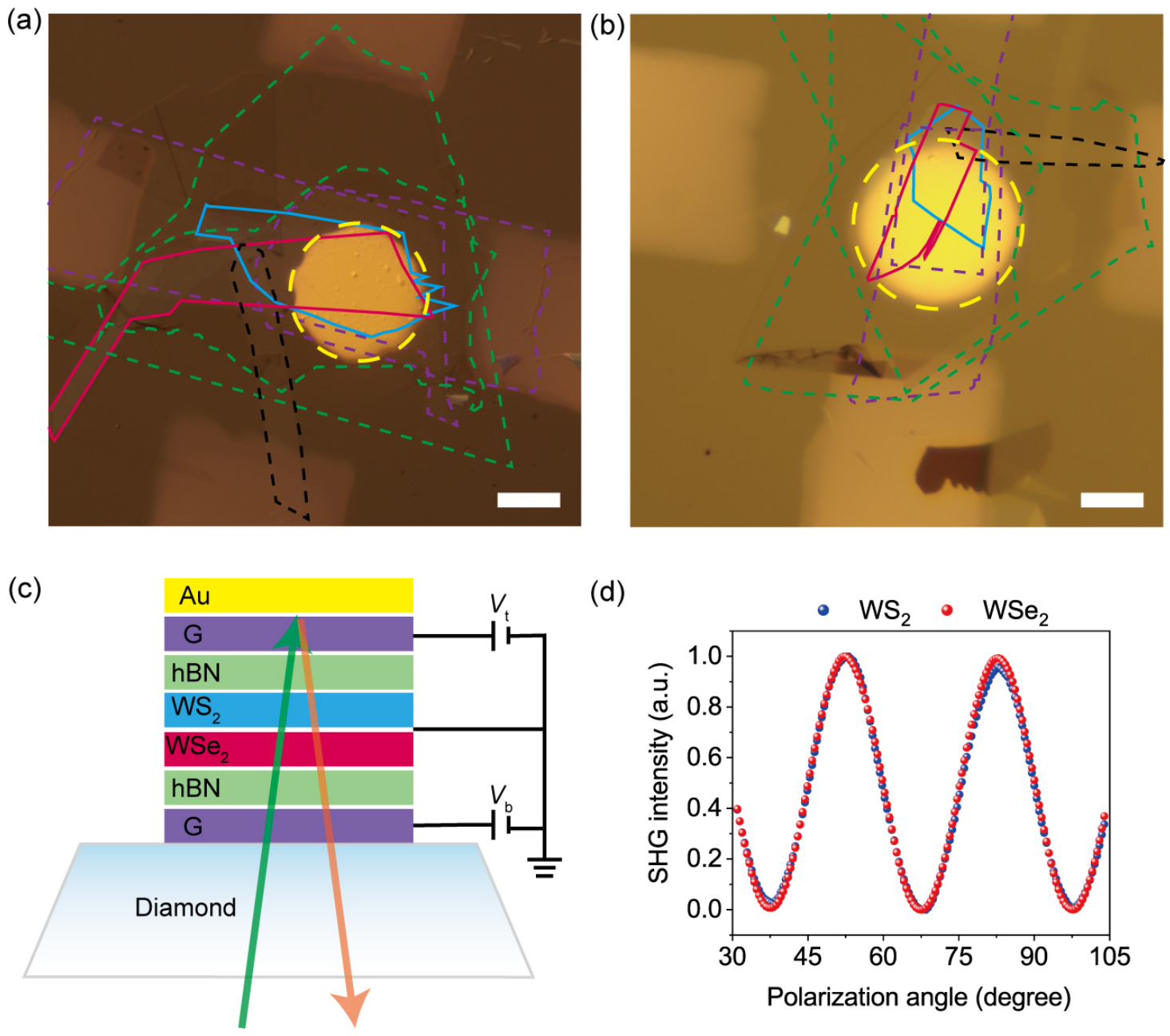

Fig. S17. Optical images of the devices and twist angle calibration. (a),(b) Optical micrographs of 60°-twisted (a) and 20°-twisted (b) bilayers residing on diamond anvil culets. The red and blue solid lines indicate the boundaries of the constituent $WSe_2$ and $WS_2$ monolayers, respectively. Scale bars, 10 μm. (c) Schematic of the dual-gated geometry. (d) Polarization dependence of SHG intensity from the individual $WSe_2$ and $WS_2$ monolayer regions of the 60°-twisted device, yielding an alignment deviation of <0.1° from 60°.

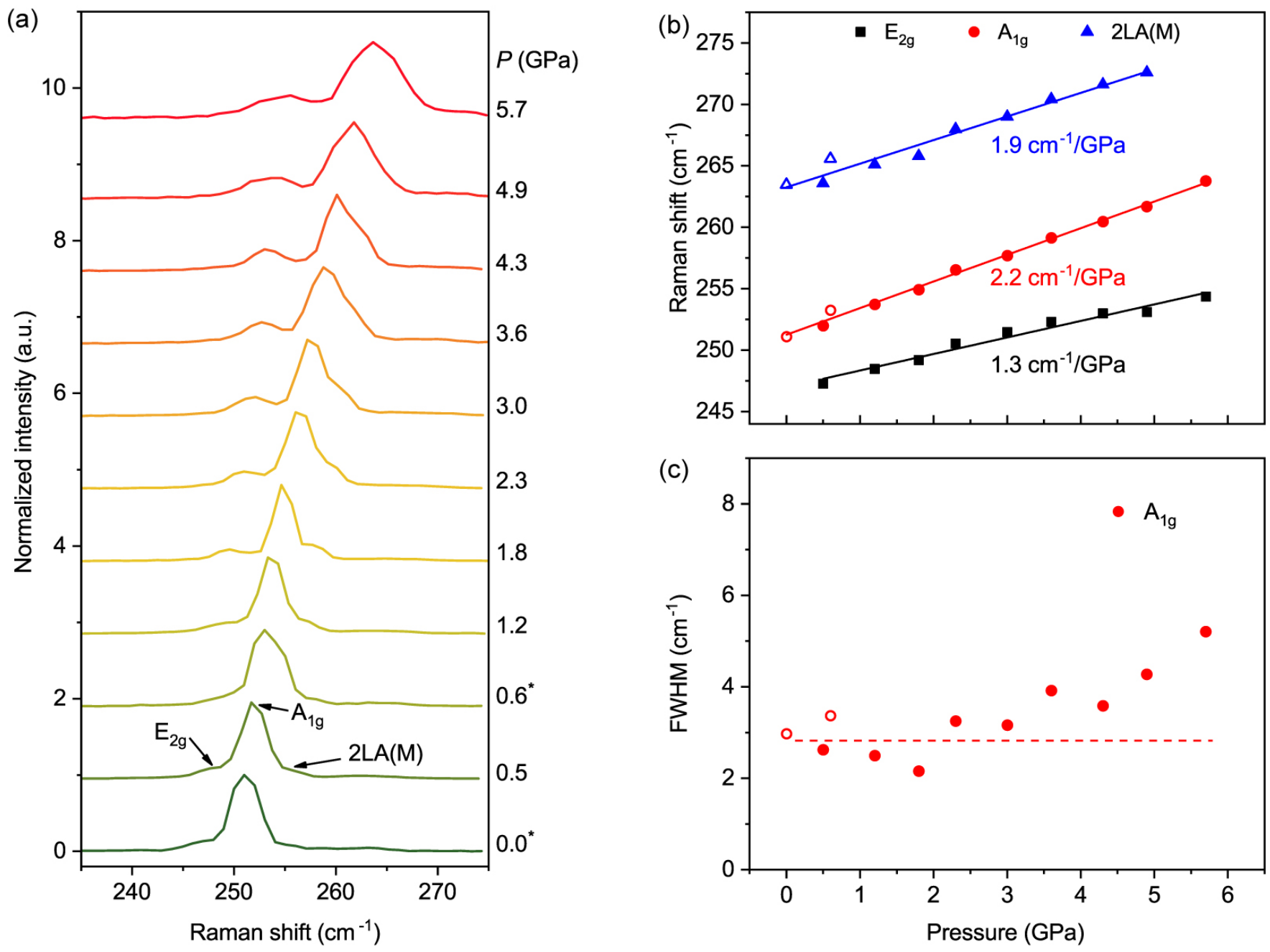

Fig. S18. Pressure evolution of phonon modes in the 60°-twisted device. (a) Pressure-dependent Raman spectra. All spectra are vertically offset for clarity. (b) Pressure dependence of the $A_{1g}$, $E_{2g}$ and 2LA phonon modes of $WSe_2$. Solid lines denote linear fits, revealing the rate of the energy shift with pressure. (c) The pressure-dependent FWHM of the $A_{1g}$ mode. The linewidth broadening above 3 GPa arises primarily from pressure-induced inhomogeneity. Data acquired during decompression are denoted either by an asterisk (a), or open symbols (b,c). The dashed line is eye-guidance (c).

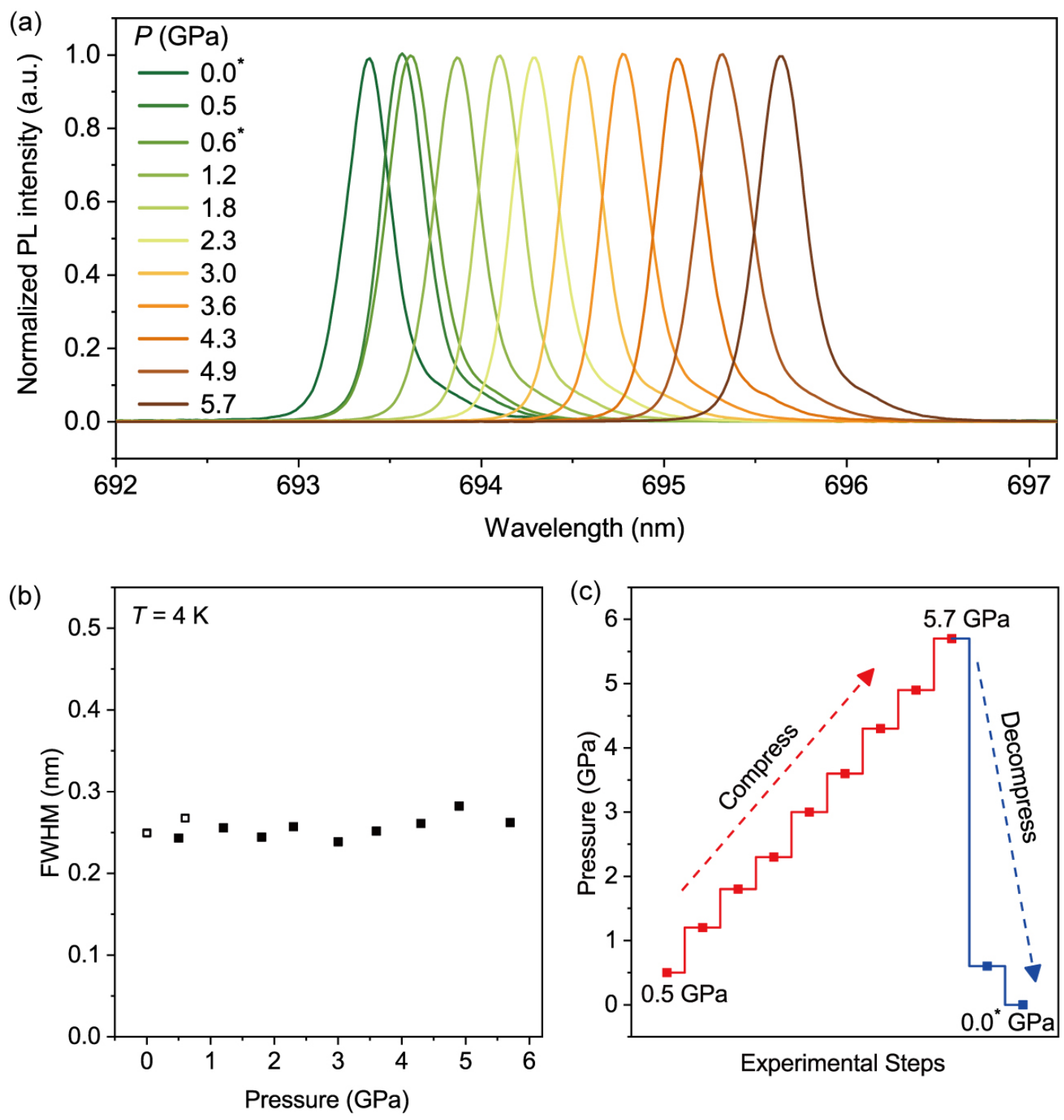

Fig. S19. Pressure calibration, hydrostaticity, and pressure cycle for the measurement cycle of the 60°-twisted device. (a) Normalized PL spectra of the ruby $R_1$ line at 4 K, utilized for in situ pressure determination. Asterisks denote data acquired during decompression. (b) The pressure-dependent FWHM of the $R_1$ peak. (c) The pressure cycle of the measurements, a compression followed by a decompression.

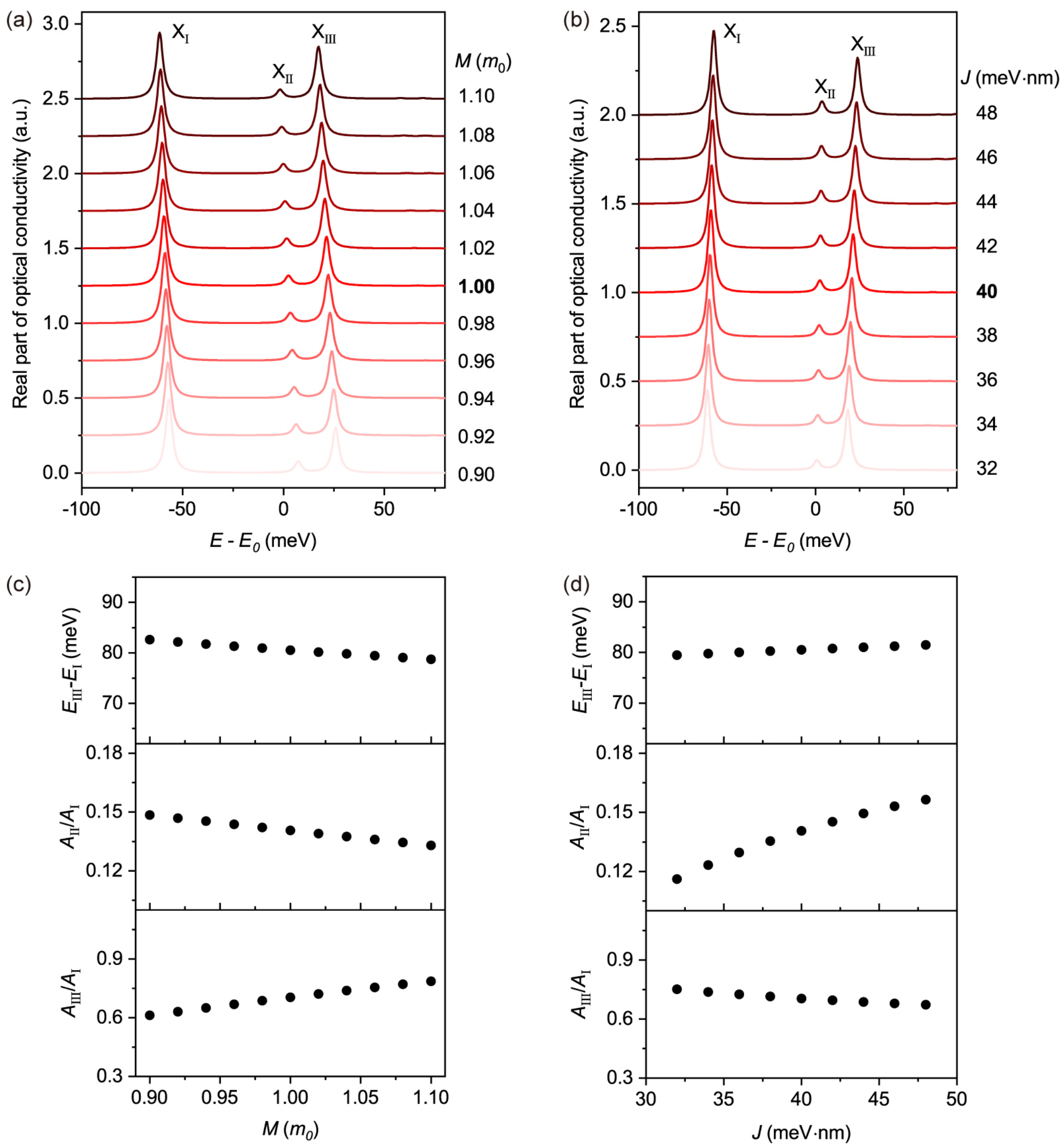

Fig. S20. Spectral responses to $M$ and $J$. (a),(b) Simulated real-part optical conductivity spectra with varying $M$ (a) and $J$ (b). Spectra are vertically offset for clarity. (c),(d) The corresponding evolution of the energy separation and relative spectral weights with varying $M$ (c) and $J$ (d). $E_i$ and $A_i$ denote the energy and spectral weight of the moiré intralayer excitons $X_i$ (i=I–III), respectively.

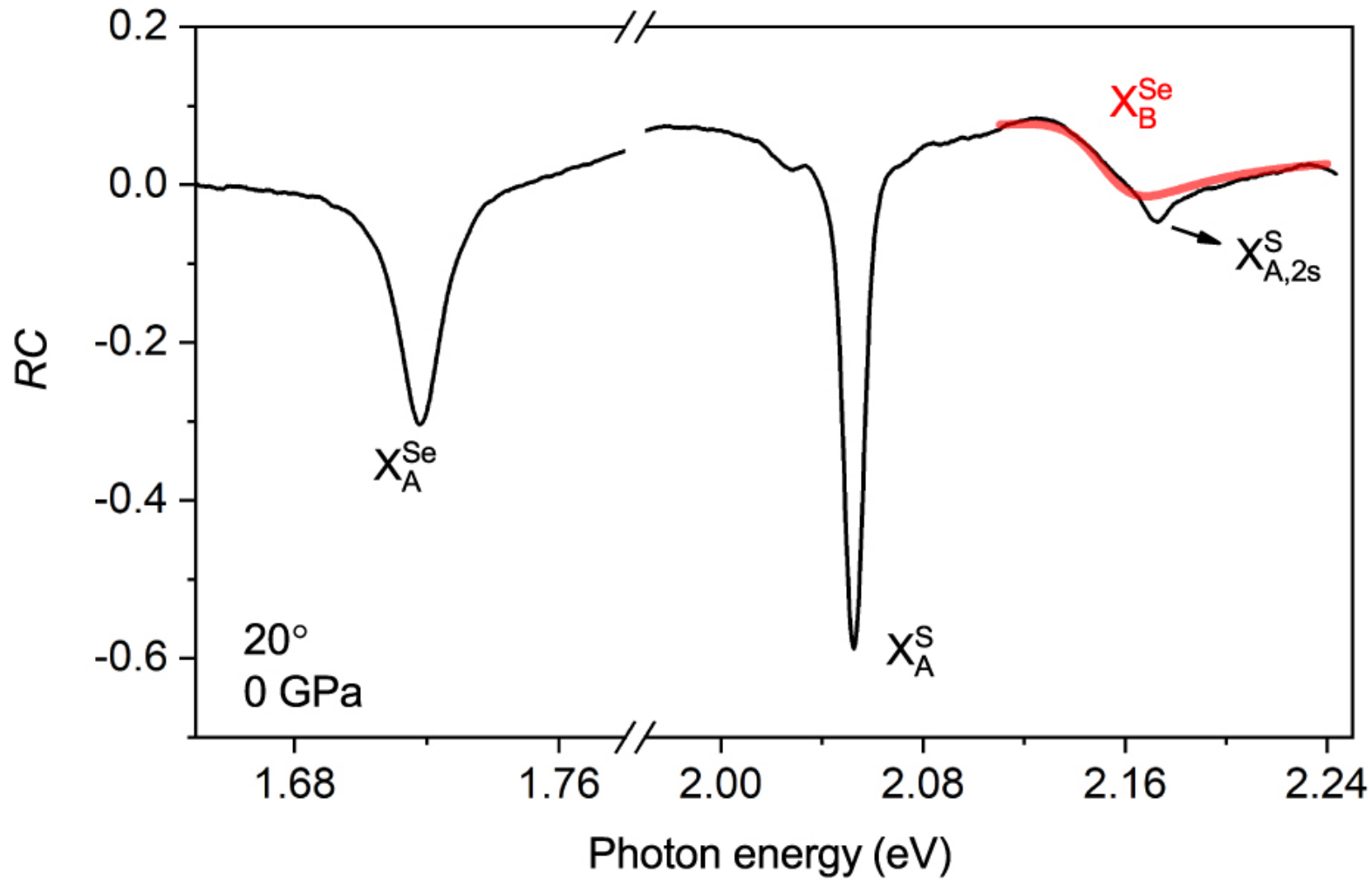

Fig. S21. Intralayer excitons in the 20°-twisted bilayer at ambient pressure. The reflectance contrast spectrum clearly exhibits the $WSe_2$ A exciton ($X_A^{Se}$) and the $WSe_2$ B exciton ($X_B^{Se}$; red curve denoting its spectral profile), as well as the $WS_2$ A exciton ($X_A^{S}$) and its 2s Rydberg state ($X_{A,2s}^{S}$). The spectral profile of $X_B^{Se}$ is determined using the phenomenological Lorentzian model described in the manuscript, yielding the energy of $X_B^{Se}$ about 2.153 eV. The energy of $X_A^{Se}$ is about 1.718 eV, giving an energy separation between $X_A^{Se}$ and $X_B^{Se}$ about 0.435 eV. For the 60°-twisted device, the energy of $X_A^{Se}$ is about 1.665 eV at ambient pressure (Fig. S6 of the manuscript), which implies the energy of $X_B^{Se}$ about 2.10 eV, assuming the same energy separation between $X_A^{Se}$ and $X_B^{Se}$.

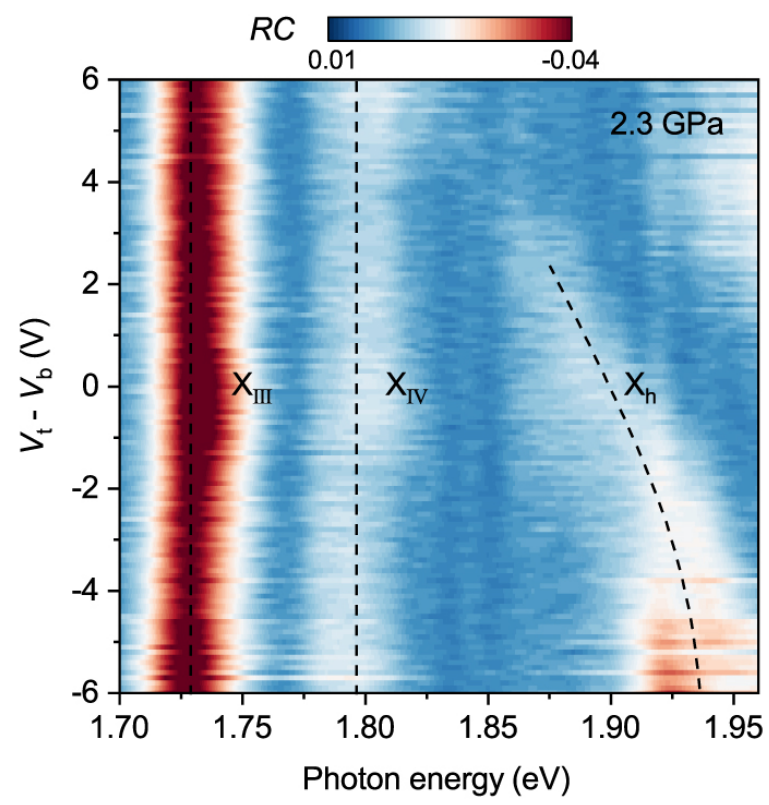

Fig. S22. The electric-field dependence of the *RC* spectrum at charge neutrality at 2.3 GPa. The dashed lines serve as guides to the eye.

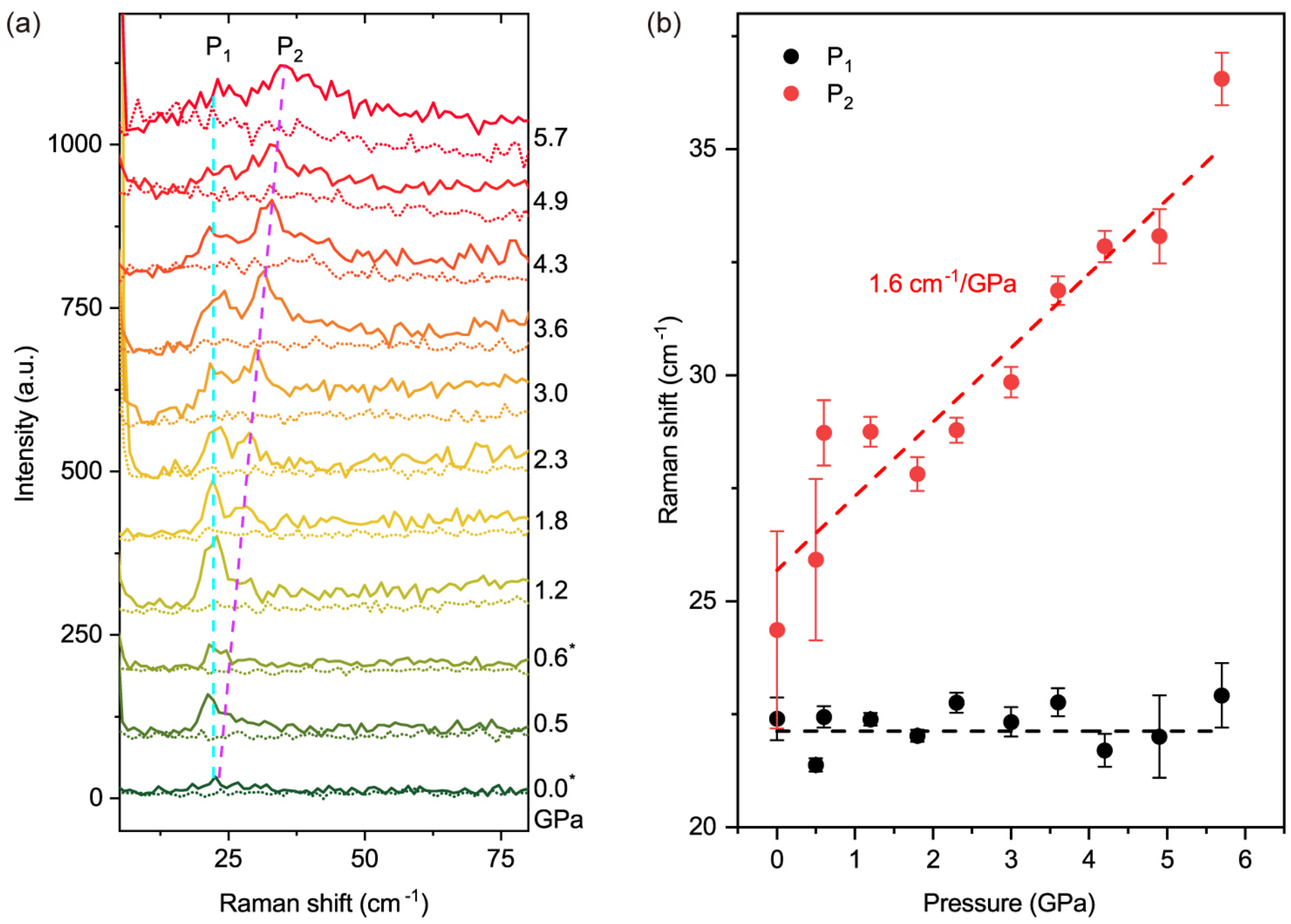

Fig. S23. Pressure dependence of low-frequency polarized Raman spectrum. (a) Pressure-dependent Raman spectra measured in the parallel-linear (XX) and perpendicular-linear (XY) polarization configurations, shown as solid and dotted curves, respectively. Spectra are vertically offset for clarity. Two modes are observed, denoted as $P_1$ and $P_2$, with their pressure evolution indicated by dashed lines. (b) Pressure dependence of the energies of $P_1$ and $P_2$. Dashed lines denote linear fits.